\documentclass[fleqn,usenatbib]{mnras}

\usepackage{newtxtext,newtxmath}
\usepackage{natbib}
\usepackage{gensymb}
\usepackage{pdflscape}	
\usepackage{soul}
\usepackage{placeins}
\usepackage{float}

\usepackage[T1]{fontenc}
\usepackage{array,longtable}
\usepackage{makecell} 
\DeclareRobustCommand{\VAN}[3]{#2}
\let\VANthebibliography\thebibliography
\def\thebibliography{\DeclareRobustCommand{\VAN}[3]{##3}\VANthebibliography}
\usepackage[pdftex]{graphicx}	
\usepackage{amsmath}
\usepackage{subcaption}

\title{Tidal features and disc thicknesses of edge-on galaxies in the SDSS Stripe\,82}

\author[Maria N. Skryabina et al.]{
Maria N. Skryabina,$^{1}$\thanks{E-mail: skryabina.mary98@gmail.com}
Kyle R. Adams,$^{2}$
and Aleksandr V. Mosenkov$^{2}$
\\
$^{1}$St.Petersburg State University, 7/9 Universitetskaya nab., St.Petersburg, 199034, Russia\\
$^{2}$Physics and Astronomy, Brigham Young University Provo, UT 84606, United States
}

\date{Accepted XXX. Received YYY; in original form ZZZ}

\pubyear{2023}

\begin{document}
\label{firstpage}
\pagerange{\pageref{firstpage}--\pageref{lastpage}}
\maketitle

\begin{abstract}
We examine deep optical images of edge-on galaxies selected from the Sloan Digital Sky Survey (SDSS) Stripe\,82. The entire sample consists of over 800 genuine edge-on galaxies with spectroscopic redshifts out to $z\sim0.2$. To discern the faintest details around the galaxies, we use three different data sources with a photometric depth of down to 30 mag\,arcsec$^{-2}$ in the $r$ band: SDSS Stripe\,82, Hyper Suprime-Cam Strategic Program, and DESI Legacy Imaging Surveys. Our analysis of the deep images reveals a variety of low surface brightness features. 49 galaxies exhibit prominent tidal structures, including tidal tails, stellar streams, bridges, and diffuse shells. Additionally, 56 galaxies demonstrate peculiar structural features such as lopsided discs, faint warps, and dim polar rings.
Overall, we detect low surface brightness structures in 94 galaxies out of 838, accounting for 11\% of the sample. Notably, the fraction of tidal structures is only 5.8\%, which is significantly lower than that obtained in modern cosmological simulations and observations. Previous studies have shown that strongly interacting galaxies have stellar discs about 1.5--2 times thicker than those without apparent interactions. In an analysis where tidal features are carefully masked for precise disc axis ratio measurements, we show that discs of galaxies with tidal features are 1.33 times thicker, on average, than control galaxies that do not have visible tidal features.
Furthermore, we find that edge-on galaxies with tidal structures tend to have a higher fraction of oval and boxy discs than galaxies without tidal features.
\end{abstract}




\section{INTRODUCTION}
\label{sec:intro}
Knowing how galaxies form and evolve over time is essential in our understanding of how the Universe appears currently. Multiple studies have found that major mergers of galaxies or minor galaxy interactions produce low surface brightness (LSB) tidal features (see e.g. \citealt{1972ApJ...178..623T}, \citealt{2000A&AS..144...85S}, \citealt{2001Natur.412...49I}, \citealt{2003ApJ...599.1082M}, \citealt{2010AJ....140..962M}, \citealt{2019MNRAS.483.1470R}). The presence of LSB structures is also confirmed in numerous simulations (e.g. \citealt{2000AJ....120.1579Y}, \citealt{2010A&A...518A..61C}, \citealt{2015MNRAS.446..521S}, \citealt{2019MNRAS.490.3196P}, \citealt{Mancillas2019}). Therefore, tracing LSB features around galaxies can help identify possible ways these galaxies have interacted and evolved over time (see e.g. \citealt{2010ApJ...715..972J}, \citealt{2015MNRAS.446..120D}, \citealt{2019MNRAS.490.1539R}, \citealt{2022RAA....22k5003M}). Specifically, in \citet{2001ApJ...557..137J}, $\Lambda$CDM cosmological galaxy formation simulations predict that from a sample of galaxies, approximately 20-40\% are expected to have tidal features at $z\sim0$. This percentage is further constrained to 25\% by \citet{Martin2022} for a similar limiting surface brightness of 28 mag\,arcsec$^{-2}$. Additionally, \citet{2022arXiv220808443V} measured a tidal feature fraction of 23\%. 
 
In the past, LSB features around galaxies were difficult to identify because they are usually too dim (fainter than $\sim25-26$~mag\,arcsec$^{-2}$ in the $r$ band) to be seen on regular photometric exposures. To add to the difficulty, some filaments of Galactic cirrus can be mistakenly identified as extragalactic LSB features \citep{2010A&A...516A..83S}. Cirrus clouds are typically found at galactic latitudes greater than 20 degrees and represent dusty filaments that do not have a particular shape. These clouds derive their name due to their similarity in appearance to the cirrus clouds of Earth. Milky Way cirrus clouds can be well-seen in deep optical images and, thus, can hinder the detection and study of LSB galaxy structures. On the other hand, deep imaging is of high importance for studying the properties of Galactic cirrus (see e.g. \citealt{2016A&A...593A...4M}, \citealt{2020A&A...644A..42R}, \citealt{2021MNRAS.508.5825M},
\citealt{2023MNRAS.519.4735S}).

Due to the difficulties discussed above and many others (e.g. sky background contamination, instrumental scattered light, observational issues related to long exposures), deep optical observations of galaxies were problematic. However, improvements in CCD and telescope technology have now made it possible to discern structures whose surface brightness reaches 29-30 mag\,arcsec$^{-2}$ (see e.g. \citealt{2010AJ....140..962M}, \citealt{2014PASP..126...55A}, \citealt{duc14}, \citealt{2019MNRAS.490.1539R}). 
New techniques in processing deep images, such as, for example, using a non-aggressive sky subtraction method \citep{2016MNRAS.456.1359F} and taking into account the point spread function (PSF, \citealt{2014ApJS..213...12J}, \citealt{2014ApJ...794..120A}, \citealt{2016ApJ...823..123T}), have become useful methods for combining images that give better depth and resolution. 
Because of the development of this field of deep photometry over the past 15-20 years, studies based on deep and ultra deep images of space objects have begun to play a relevant role in the study of the Universe (see e.g. \citealt{2005ApJ...631L..41M}, \citealt{2013ApJS..209....6I}, \citealt{2014PASP..126...55A}, \citealt{2015ApJ...798L..45V}, \citealt{2015ApJ...807L...2K}, \citealt{2021MNRAS.503.6059P}).
 
Modern sky surveys, such as, for example, the Sloan Digital Sky Survey (SDSS, \citealt{2000AJ....120.1579Y}) and the much deeper DESI Legacy Imaging Surveys \citep{2019AJ....157..168D} have provided a plethora of data for extragalactic studies. The derivation of such a large amount of observational data has created a huge scope for new studies of objects that were previously inaccessible for observations. The upcoming Legacy Survey of Space and Time (LSST, \citealt{2019ApJ...873..111I}) will further enhance our ability to image galaxies, promising even deeper and more comprehensive observations that could unveil extremely faint extragalactic structures.

In particular, one region of the sky most interesting for studying is the SDSS Stripe\,82 \citep{Jiang_2014}. This stripe has a width of 2.5 degrees along the celestial equator with the following coordinates: -50$\degree$$<$R.A.$<$ 60$\degree$, -1.25$\degree$$<$Dec.$<$1.25$\degree$. The total area of the stripe is 275 square degrees in all five SDSS bands. Specifically for our study, the SDSS Stripe\,82 has an advantage in its depth. For instance, ordinary SDSS exposures reach a depth of 26.5 mag\,arcsec$^{-2}$ in the SDSS $r$ band, but in the Stripe\,82 the average value is 28.6 mag\,arcsec$^{-2}$, as provided by the IAC Stripe\,82 Legacy Project\footnote{\url{http://research.iac.es/proyecto/stripe82/}}. The photometric depth of images in the SDSS Stripe\,82, as we show in this paper, is sufficient to explore the prominent low-surface brightness structures around galaxies (see also \citealt{2016MNRAS.456.1359F}, \citealt{2019A&A...629A..12M}, \citealt{2021ApJ...923..205Y}, \citealt{2021ApJS..257...60Z}). 

Optical observations of edge-on galaxies provide an unprecedented view of the luminous matter distribution in galaxies (\citealt{1981A&A....95..105V}, \citealt{1998MNRAS.299..595D}, \citealt{2002AJ....124.1328D}, \citealt{2002MNRAS.334..646K}, \citealt{2010MNRAS.401..559M}, \citealt{2014ApJ...787...24B}). This view allows LSB features to be seen with increased visual clarity, especially above and below the galaxy mid-plane, unlike face-on galaxies where faint structures are much less apparent over the galaxy body and only become discernible far beyond the optical radius  (\citealt{1999AJ....118.1209F}, \citealt{2019ApJ...883L..32V}, \citealt{2020MNRAS.494.1751M}, \citealt{2020ApJ...897..108G}, \citealt{2020MNRAS.497.2039M}, \citealt{2021MNRAS.506.5030M}, \citealt{2022ApJ...932...44G}, \citealt{2022MNRAS.515.5698M}). 

Using machine learning or automated parametric or non-parametric methods is helpful in identifying the morphology of galaxies (e.g. \citealt{2018ApJ...866..103K}, \citealt{2022A&A...662A.124S}) and possibly tidal features (\citealt{2023MNRAS.521.3861D}). In addition, neural networks have identified cirrus clouds in deep optical images, specifically, in Stripe\,82 \citep{2023MNRAS.519.4735S}. When it comes to galaxy LSB structures themselves, machine learning has not yet been trained to identify and separate them from galactic cirrus, extended halos of bright, saturated stars, and image artefacts (especially, artefacts of sky background subtraction). Also, our training sample of LSB features is still insufficiently large for developing such an automated classification. Even with automated classification methods, studies still use human identification to ensure avoiding improper classifications (\citealt{2020MNRAS.498.2138B, 2023A&A...671A.141M, 2024A&A...681L..15M}).

In this study, we create a new catalogue of edge-on galaxies selected from the deep Stripe\,82, ES82 for short. This work is aimed at identifying and studying LSB features around the selected {\it edge-on} galaxies while taking into account cirrus contamination based on the recent study by \citet{2023MNRAS.519.4735S} and by using complementary sources of deep imaging to exclude false identifications. We look at the percentage of tidal features from our catalogue to compare these statistics with those in recent papers, where galaxies at random orientations were explored, and with modern cosmological hydrodynamical simulations. 

In numerous studies, it has been found that galaxy interactions cause a thickening of the disc (see 
 e.g. \citealt{1988ApJ...331...71S}, \citealt{1993ApJ...403...74Q}, \citealt{1996ApJ...460..121W}, \citealt{2003ApJ...597...21A}, \citealt{2011A&A...530A..10Q}, \citealt{2008MNRAS.391.1806V}, \citealt{2009MNRAS.399..166V}). In \citet{1997A&A...324...80R}, it is shown by the ratio of the radial and vertical scales $h_\mathrm{R}/z_0$ that interacting galaxies, on average, have about 1.5-2 times thicker discs than non-interacting galaxies \citep[see also][]{Schwarzkopf2001}. This fact can be explained by the influence of the gravitational forces exerted during interactions with other galaxies. Such interactions can lead to significant disturbances in a galaxies stellar disc, such as warping, heating, or flaring of the disc material \citep{2009ApJ...707L...1B,2011A&A...530A..10Q,2011ApJ...741...28C,2016MNRAS.461.4233R}. These disturbances increase the velocity dispersion among the stars in the disc, essentially stirring them up, and causing  the disc to thicken as stars move further from the galactic plane.
Also, as recently found by \citet{Kumar2021} through N-body simulations of minor galaxy flybys (which can create warps, stellar streams, and tidal bridges), such fast encounters can increase the disc scale height $z_0$ by only a maximum of $\sim$ 4\%, but the reduction of the disc scale length $h_\mathrm{R}$ leads to a thickening of the stellar disc. The relationship between interactions, which produce tidal structures and a thickened disc, is still not well studied. Therefore, in our work, our secondary aim is to shed more light on this issue. To this end, we compare the mean value of the galaxy disc thicknesses for two samples of galaxies: a sample of objects with LSB features and those without them.

This paper is organised as follows. In Section~\ref{sec:sample}, we discuss the sample selection, describe the process of image preparation, determine the inclination of galaxies in the sample, and the sample completeness is calculated. In Section~\ref{sec:classification}, we provide definitions for each type of tidal structure and prominent structural feature, accompanied by examples from our sample and statistics. In Section~\ref{sec:decomp}, the method of photometric decomposition is outlined and the results on the disc thickness and diskyness/boxyness $C_0$ parameter are presented. In Section~\ref{sec:discussion}, we discuss the results obtained. We summarise our findings in Section~\ref{sec:summary}.

\section{The sample and data}
\label{sec:sample}
There have been numerous projects dedicated to processing or reprocessing SDSS Stripe\,82 images. One such project is the IAC Stripe 82 Legacy Project \citep{2016MNRAS.456.1359F}. This project employs a non-aggressive technique for sky background subtraction to preserve low-surface brightness structures across different spatial and intensity scales. We have leveraged data from the IAC Stripe 82 Legacy Project to develop a comprehensive catalogue of edge-on galaxies within the deep Stripe\,82 region, referred to as ES82 (Edge-on galaxies in SDSS Stripe 82). This catalogue provides an excellent opportunity for further analysis of these galaxies, enabling the detection and characterisation of low-surface brightness features such as tidal tales, stellar streams, arcs, diffuse shells, bridges, and disc deformations. These structures serve as compelling evidence of galaxy interactions with their surrounding environments.

\subsection{Selection of galaxies}

To create our catalogue of edge-on galaxies, we used several sources to identify flat galaxies in Stripe\,82. First, we utilised existing {\sc SExtractor} \citep{1996A&AS..117..393B} catalogues of objects in the $r$ band provided by \citet{2016MNRAS.456.1359F} for this sky region. Our aim is to select galaxies with a relatively large angular size and small apparent axis ratio. Since {\sc SExtractor} does not list the optical diameter of galaxies, we used the Kron radius ($r_{Kron}$) for size filtering. Our selection criteria, based on the analysis results from the EGIS and EGIPS catalogues (see \citealt{2014ApJ...787...24B} and \citealt{2022MNRAS.511.3063M}, respectively), include a Kron radius ($r_{Kron}$) greater than 15~arcsec and an {\sc SExtractor} axis ratio B\_IMAGE/A\_IMAGE of less than 0.3. These simple criteria should help identify sufficiently large edge-on or nearly edge-on galaxies, enabling detailed resolution of their vertical structures.

\begin{figure}
\includegraphics[width=8cm]{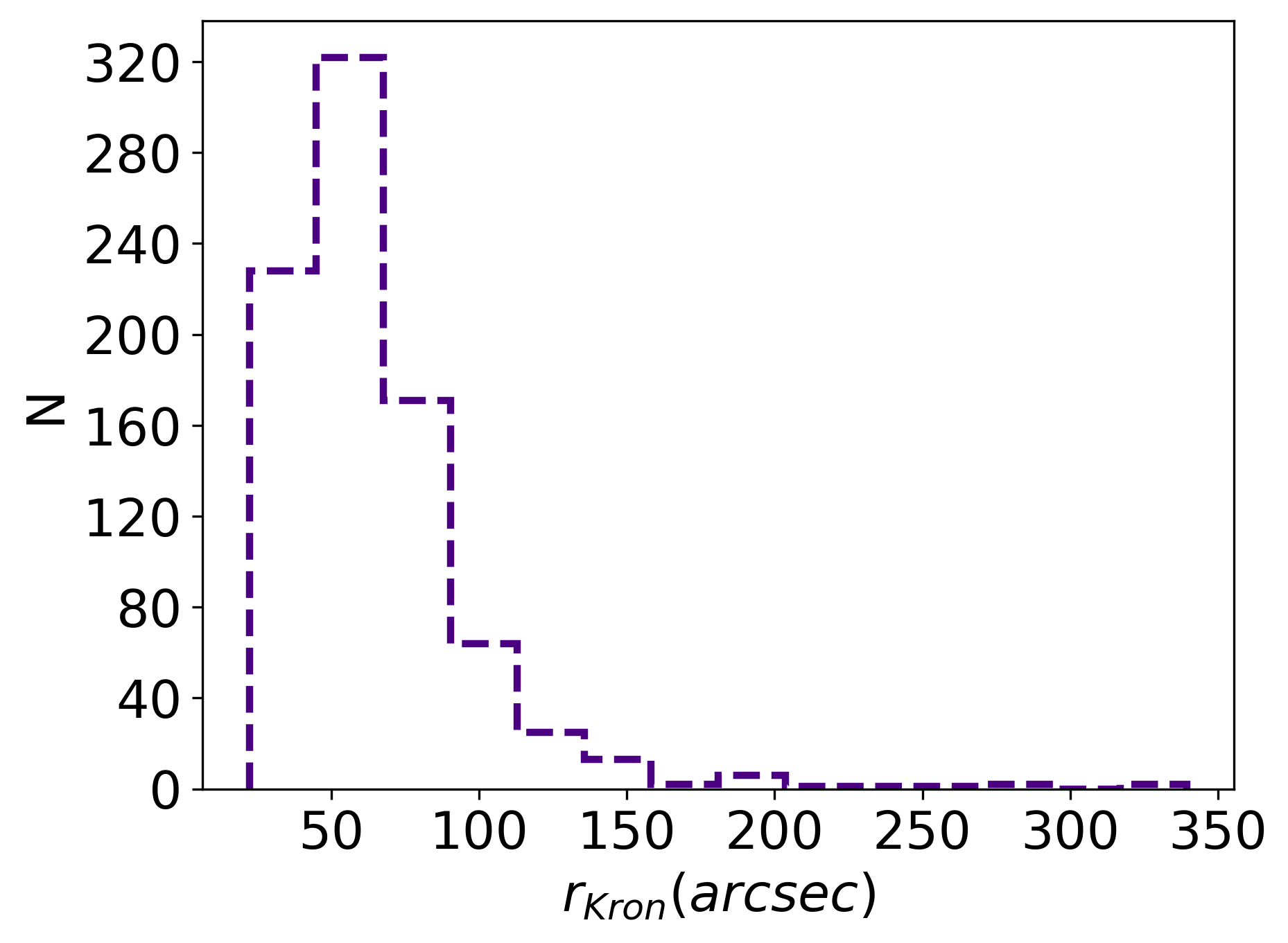}\\
\caption{Kron radius distribution for galaxies from the catalogue.}
\label{fig:r_kron}
\end{figure}

Additionally, we referred to the EGIS \citep{2014ApJ...787...24B} and Galaxy Zoo \citep{2016MNRAS.461.3663H} catalogues (for the latter, we considered the fraction of votes for the ``edge-on'' response to be >80\%) to ensure that all edge-on galaxies are selected according to our automated criteria. Finally, we visually examined the catalogue of $\sim$17,000 galaxies in Stripe\,82 from \citet{Bottrell_2019} to identify edge-on galaxies and eliminate duplicates from the pre-final sample by comparing the coordinates of all objects among themselves.

Next, we used the SDSS Stripe\,82 \citep{2016MNRAS.456.1359F}, the Hyper Suprime-Cam Subaru Strategic Program (HSC-SSP) PDR3 \citep{2022PASJ...74..247A}, and DESI Legacy Imaging Surveys \citep{2019AJ....157..168D} to create RGB images for each selected galaxy using these three different sources (the image creation process is discussed in detail in Sect.~\ref{sec:data}). Note that SDSS Stripe\,82 is not completely covered by the HSC-SSP and DESI Legacy surveys, as also described in Sect.~\ref{sec:data}. By visual inspection of these images for all the selected galaxies, we eliminated objects that did not appear to be real edge-on galaxies due to the presence of image artefacts that resembled an edge-on disc, spikes of very bright stars, or regions of spiral structure. We also removed galaxies with inclinations that are not sufficiently large (i.e., $i\lesssim80\degree$), as evidenced by visible spiral arms, a discernible stellar disc plane, or noticeably shifted dust lanes. After conducting a thorough visual inspection of the prepared RGB images for a sample of 1167 galaxies, we selected the final version of the catalogue consisting of 838 genuine edge-on or nearly edge-on galaxies the inclinations of these galaxies are considered in Sect.~\ref{sec:inclinations}). The size distribution for the catalogue galaxies is presented in Fig.~\ref{fig:r_kron}.

\subsection{Data and Preparation}
\label{sec:data}

All fits images of the selected galaxies are taken from the IAC Stripe\,82 Legacy Project\footnote{\url{http://research.iac.es/proyecto/stripe82/}}  \citep{2016MNRAS.456.1359F}. Specifically, we use sky-rectified images in the $g$, $r$, $i$, and $r$-deep wavebands, along with the corresponding empirical Point Spread Function (PSF) images and weight images. These are utilised in the galaxy photometric decomposition described in Sect.~\ref{sec:decomp}. For the coadded data set, the mean surface brightness limit reaches $\mu$[3$\sigma$, $10 \times$ 10 arcsec$^2$] = 28.6 mag\,arcsec$^{-2}$ for the $r$ band\footnote{Using the 3$\sigma$ level in square boxes of $10\times10$~arcsec$^{2}$ has become one of the standard definitions for photometric depth, facilitating comparisons between results obtained with different telescopes and instruments.} The median value of the FWHM for the same $r$ band is 1.10~arcsec, with a pixel size of 0.396~arcsec.

To enhance the classification process of the sample galaxies, we utilise the most recent releases of two additional deep sky surveys: DESI Legacy Imaging Surveys \citep{2019AJ....157..168D} and the HSC-SSP \citep{2022PASJ...74..247A}. 

The HSC-SSP survey utilises the Hyper Suprime-Cam (HSC) on the 8.2m Subaru Telescope, featuring three layers: wide (1400 deg$^2$, $5\sigma$ limit $r\sim26$~mag), deep (27 deg$^2$,  $5\sigma$ limit $r\sim27$~mag), and ultradeep (3.5 deg$^2$,  $5\sigma$ limit $r\sim28$~mag). The HSC employs 104 science CCDs and covers a 1.5-degree diameter field of view with a pixel scale of 0.168~arcsec. Notably, the coadded data from the HSC-SSP offers improved resolution, enhancing the visualisation of low surface brightness structures, with an average surface brightness depth of $29.6\pm0.4$~mag\,arcsec$^{-2}$ in the $r$ band.

The DESI Legacy Imaging Surveys, combining the Dark Energy Camera Legacy Survey (DECaLS), the Beijing-Arisona Sky Survey (BASS), and the Mayall $z$-band Legacy Survey (MzLS), cover approximately 14,000 deg$^2$ of the extragalactic sky across both the Northern and Southern celestial hemispheres in the $g$, $r$, and $z$ bands. The average surface brightness depth in the $r$ band for our sample galaxies is $28.4\pm0.3$~mag\,arcsec$^{-2}$.


As the observed fields overlap partially in the IAC Stripe 82, DESI Legacy, and HSC-SSP, not all galaxies in SDSS Stripe\,82 have observations in all three surveys: 539 out of 838 galaxies in our catalogue have HSC-SSP images and 827 galaxies have data in DESI Legacy. This means that for at least 65\% of the galaxy sample, we have three different data sources for more reliable structure classification. 

For a quantitative analysis of galaxy images, we use the SDSS Stripe\,82 images, which have been specially prepared by the IAC Stripe\,82 Legacy Project for the accurate treatment of low-surface brightness structures. To enhance galaxy images, we employ a semi-automatic image preparation pipeline (IMage ANalysis or {\sc IMAN})\footnote{\url{https://bitbucket.org/mosenkov/iman_new/src/master/}} and additional Python scripts. Below, we briefly describe our methodology for preparing stacked cut-outs of the sample galaxies. 

First, we make initial image cropping and rotation using the galaxy centre coordinates and position angles from the {\sc SExtractor} catalogue files provided in \citet{2016MNRAS.456.1359F}. After visual inspection, position angles for about 40\% of the galaxies are corrected by inscribing an ellipse to fit the outer isophotes of the galaxy and determining its position angle. After that, the first step with cropping and rotation is performed again. By final cropping, we fix the size of each image as a square with sides equal to 6 semi-major galaxy axes. The rotation and cropping procedures are also applied to the error maps which contain information about the weight of each pixel as a measure of the photometric quality. Employing the Python package \textit{photutils}\footnote{\url{https://photutils.readthedocs.io/en/stable/}}, masks for all objects that do not belong to the target galaxy are created. Sky subtraction is not executed as our images have already been sky-rectified. We carry out the procedures, described above, for all bands to obtain cropped and rotated coadds (using the deep $r$-band images) and colour RGB images to exploit the Stripe\,82 data to its fullest. This aids in unveiling low-surface brightness structures around the galaxies. The mosaic of SDSS coadded images of the edge-on disc sample is available online \footnote{\url{https://physics.byu.edu/faculty/mosenkov/docs/Edge-on_Stripe82.pdf}}. 
Supplementary material also contains additional imagery of the galaxies sourced from the different surveys.

\subsection{Inclinations of the selected galaxies}
\label{sec:inclinations}
Determining the exact orientation of a disc galaxy can often be problematic, especially for angularly small galaxies where we cannot clearly see the orientation of the dust lane. Therefore, we classify the selected galaxies into the following groups, which we will use in further analysis:

\begin{enumerate}
    \item True edge-on galaxies. This group includes objects that have clear, well-resolved signs indicating the inclination of the object. These signs include the presence of a dust lane that dissects the galaxy body into two approximately equal parts under and above the dust lane, the shape of the disc isophotes (round isophotes often signify that the galaxy is not seen purely edge-on \citealt{2020MNRAS.494.1751M}), the absence of visible in-plane rings, spiral arms, and other global non-axially symmetric components such as a bar \citep{2020MNRAS.497.2039M}. In this subsample, we do not include pure edge-on galaxies that exhibit prominent structural features in or around the galaxy that would affect the quality of our photometric fitting (see Sect.~\ref{sec:decomp}) and determination of the disc thickening (all such galaxies are moved to the fourth group described below). In total, this subsample comprises 242 galaxies.
    \item The second group consists of galaxies without obvious signs of their inclination. These objects are presumably visible to us from the edge, but there are no clear signs of this. In particular, these are angularly small objects, for which the resolution of our images (even in the HSC-SSP where the PSF FWHM is the best among the surveys used, about 0.6~arcsec) is not sufficient to determine the galaxy inclination. In this group, we have 409 objects.
    \item The third group contains objects whose orientation is close to edge-on. For them, the inclination angle is most likely $80\degree<i<85\degree$. We can suppose this by the shifted central dust lane relative to the galaxy plane (see \citealt{2015MNRAS.451.2376M}) or the barely visible ends of the spiral arms. This sample consists of 73 objects.
    \item The fourth group contains objects that did not fall into the first group: they are seen edge-on but demonstrate prominent structural and tidal features,  that can affect the determination of the disc thickening. In this group, we also include galaxies in the vicinity of which there are very bright sources that can outshine the target galaxy. This sample comprises 115 objects.
\end{enumerate}

The main reason for creating this classification is to compare the thickness of the galaxy discs, and take into account the inclination effects which is important for making reliable conclusions. To determine the apparent axis ratio $q$ more accurately, we perform a simple S\'ersic photometric decomposition described in Sect.~\ref{sec:decomp}.

\subsection{Sample Completeness and the Catalogue}

To evaluate the sample's completeness, we employ a straightforward $V/V_m$ test outlined in \citet{1979ApJ...231..680T}. This test involves calculating the volume of a sphere ($V$) with a radius equivalent to the distance to the galaxy ($D$). The formula for $D$ is given by $D = d/\theta$, where $d$ represents the linear diameter of the galaxy and $\theta$ denotes the angular diameter. Additionally, we determine the volume of another sphere ($V_m$), which has a radius equal to the maximum distance ($D_m$) at which the galaxy can still be considered part of our sample. In our scenario, $D_m = d/\theta_L$, where $\theta_L$ signifies the smallest angular size observed in our sample object. For our calculations, we use $\theta_L = 22.33$ arcsec, considering the angular size as the value of the galaxy's Kron radius. The volume ratio, denoted as $V/V_m$, is given by $(\theta_L/\theta)^3$. If objects in Euclidean space are distributed uniformly and the sample is complete, then the volume ratio ($\langle V/V_m\rangle$) will be equal to 0.5.

Calculating this $V/V_m$ value for the catalogue of objects, successively excluding objects with small angular sizes, we conclude that for objects with $r_{Kron}$ > 61 ~ arcsec (45\% of the sample), our sample is essentially complete with an average volume ratio of $\langle V/V_m\rangle=0.49$ in the $r$ waveband.

\begin{figure}
\includegraphics[width=8cm]{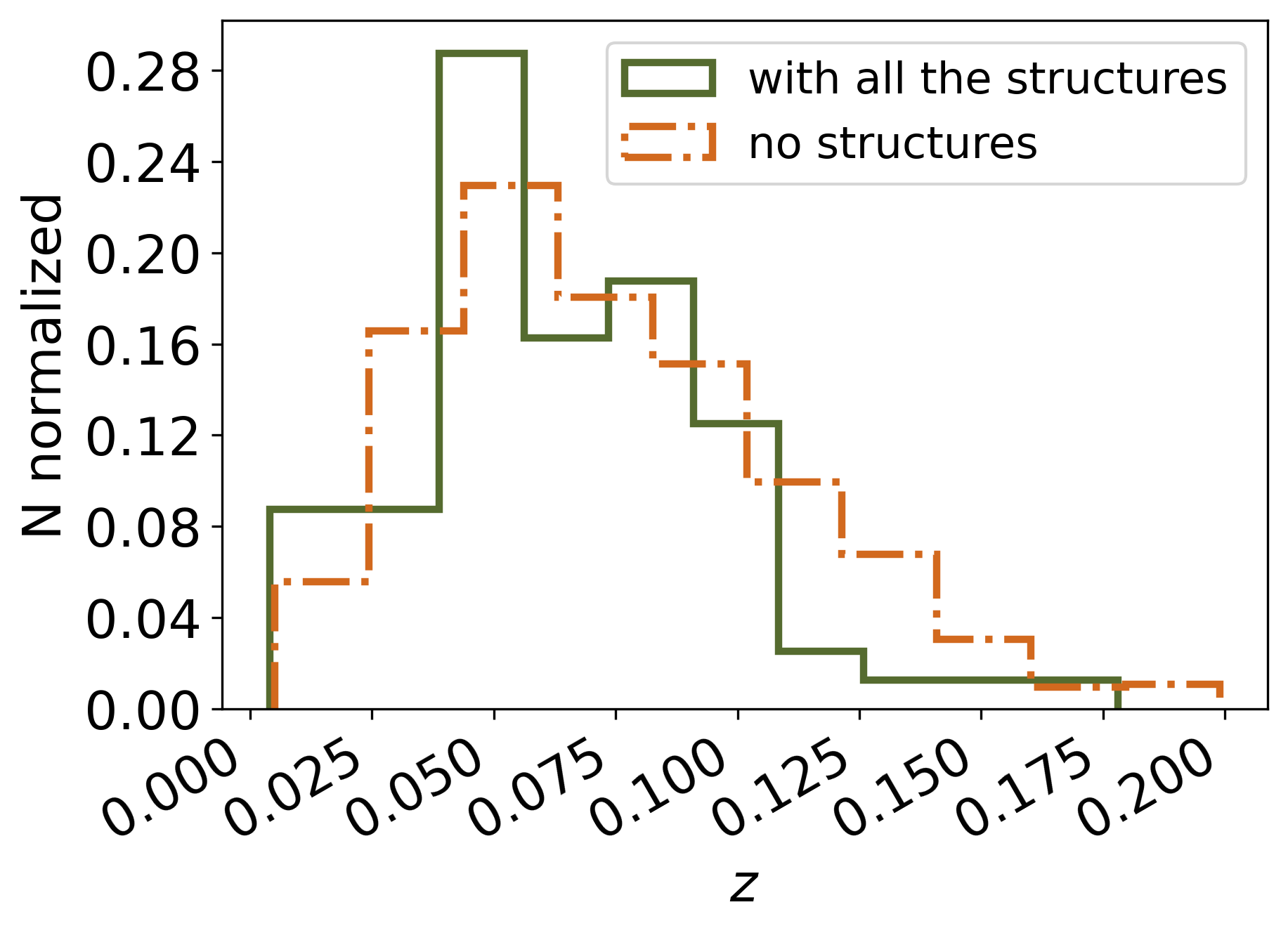}\\
\caption{Redshift distribution for galaxies from the catalogue.}
\label{fig:z_spec}
\end{figure}

The whole catalogue is available in the online journal. The catalogue’s core table is our Table \ref{table:table3} (with R.A., Dec., PA, $q$, $C_0$, redshift, the total magnitude in the $r$ band, etc.). In Fig.~\ref{fig:z_spec} we show the distribution of our galaxies by redshift, with and without prominent structural features. As one can see, both subsamples have very similar distributions which peak at $z\simeq0.06$. The quartiles for both subsamples are 0.06$^{+0.03}_{-0.02}$. The galaxies in our catalogue span a typical range of absolute magnitudes, with a median value of -20.7$^{+0.8}_{-0.7}$ (see Fig.~\ref{fig:z_spec}).

\begin{figure}

\includegraphics[width=8cm]{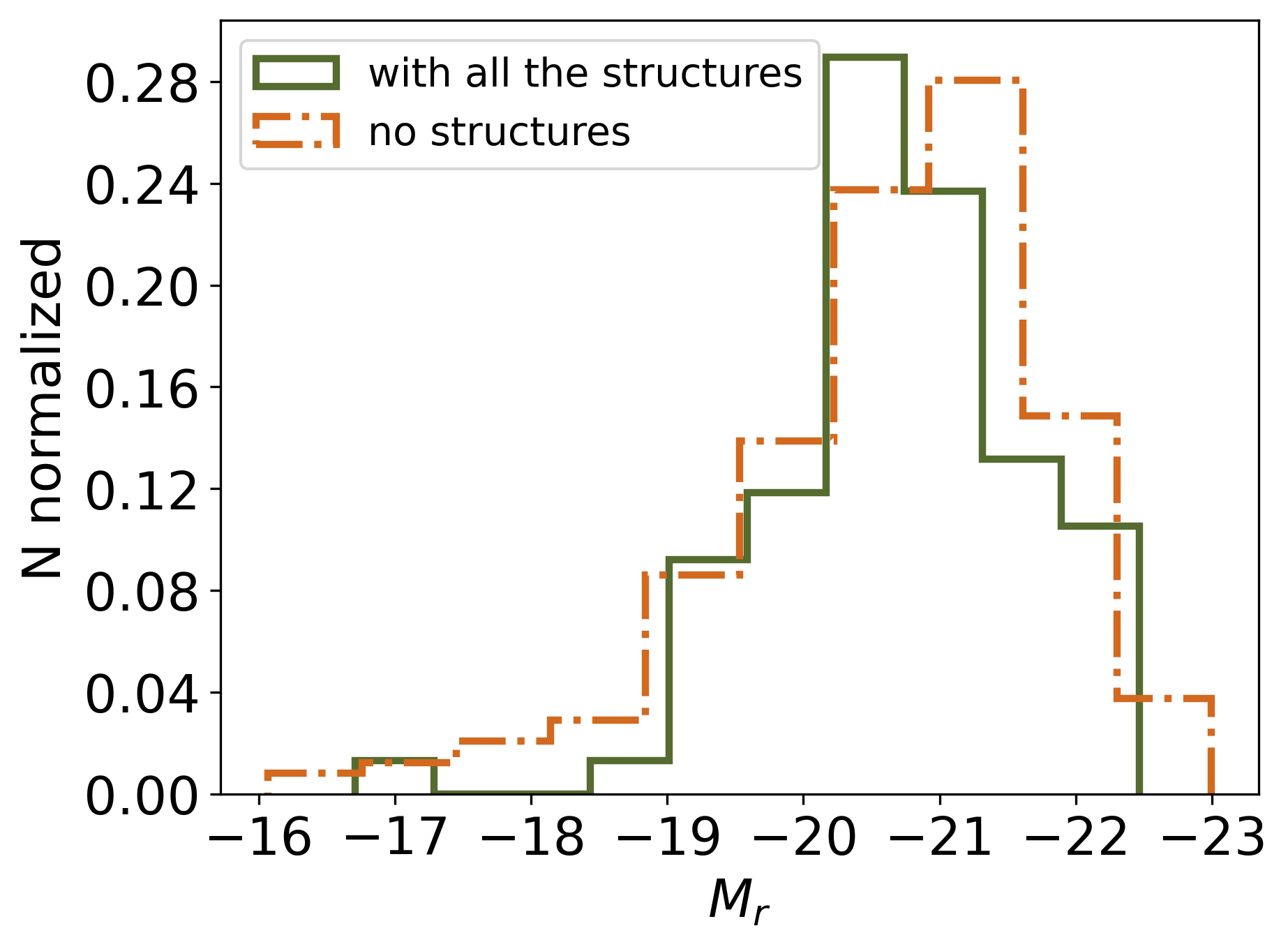}\\
\caption{Absolute magnitude distribution for galaxies from the catalogue.}
\label{fig:M_abs}
\end{figure}

\section{STRUCTURE CLASSIFICATION AND STATISTICS}
\label{sec:classification}

We carry out an LSB structure classification by visual inspection of images from the three independent deep surveys. Using four images, namely the stacked and colour RGB images based on the SDSS Stripe\,82, DESI Legacy, and HSC-SSP data, we reveal low surface brightness structures near some galaxies from the catalogue. All the structures are divided into two categories: tidal structures and distinctive structural features. In the first category, we include:

\begin{enumerate}
    \item Tidally Distorted Satellites or Satellite Debris: The remnants of satellites that were disrupted in the process of tidal interaction (see \citealt{1998ApJ...508L..55M}, \citealt{1999AJ....118.1709M}, \citealt{2001ApJ...557..137J}, \citealt{2018MNRAS.478.3879S}, \citealt{2012Natur.482..192R}).
    \item Tidal Tails and Stellar Streams: Elongated structures of stars and interstellar gas (manifesting itself in the optical through star-forming regions) that extend into space from a galaxy due to tidal interactions. Stellar streams are created by the tidal disruption of low-mass satellites interacting with the host galaxy (\citealt{1996ApJ...465..278J}, \citealt{1999AJ....118.1709M}, \citealt{McConnachie_2009}).
    Tidal tails result from the merging and interaction of galaxies, typically through major mergers \citep{1992AJ....103.1089B,2013LNP...861..327D,2015ApJ...807...73O}.
    \item Diffuse Shells, Plumes, or Asymmetric Stellar Halos: These diffuse or concenctric ring-like structures can be produced by major mergers or multiple minor merger events. They are often associated with the accretion of satellites on nearly radial orbits (\citealt{1990dig..book...72P}, \citealt{2011ApJ...743L..21C}, \citealt{2013arXiv1312.1643E}, \citealt{2018MNRAS.480.1715P}).
    \item Bridges: Extended structures consisting of stars and gas representing the flow of matter from one object to another \citep{1972ApJ...178..623T,2010AJ....139.1212S,2017MNRAS.472.2554B}.
    \item Single Arcs and Loops: Elongated structures around galaxies also produced by tidal effects \citep[see e.g.][and references therein]{2018A&A...614A.143M}, specifically by dynamical friction \citep{2008MNRAS.383...93B} and tidal stripping \citep{2008ApJ...673..226P}.
    \item Disc Deformations: These include thick tidal warps \citep{2020MNRAS.498.3535S} and deformed or distorted stellar discs \citep{2022MNRAS.513.1867D,2023MNRAS.518.2870D}.
\end{enumerate}

\begin{figure*}
\centering
\includegraphics[width=13.5cm]{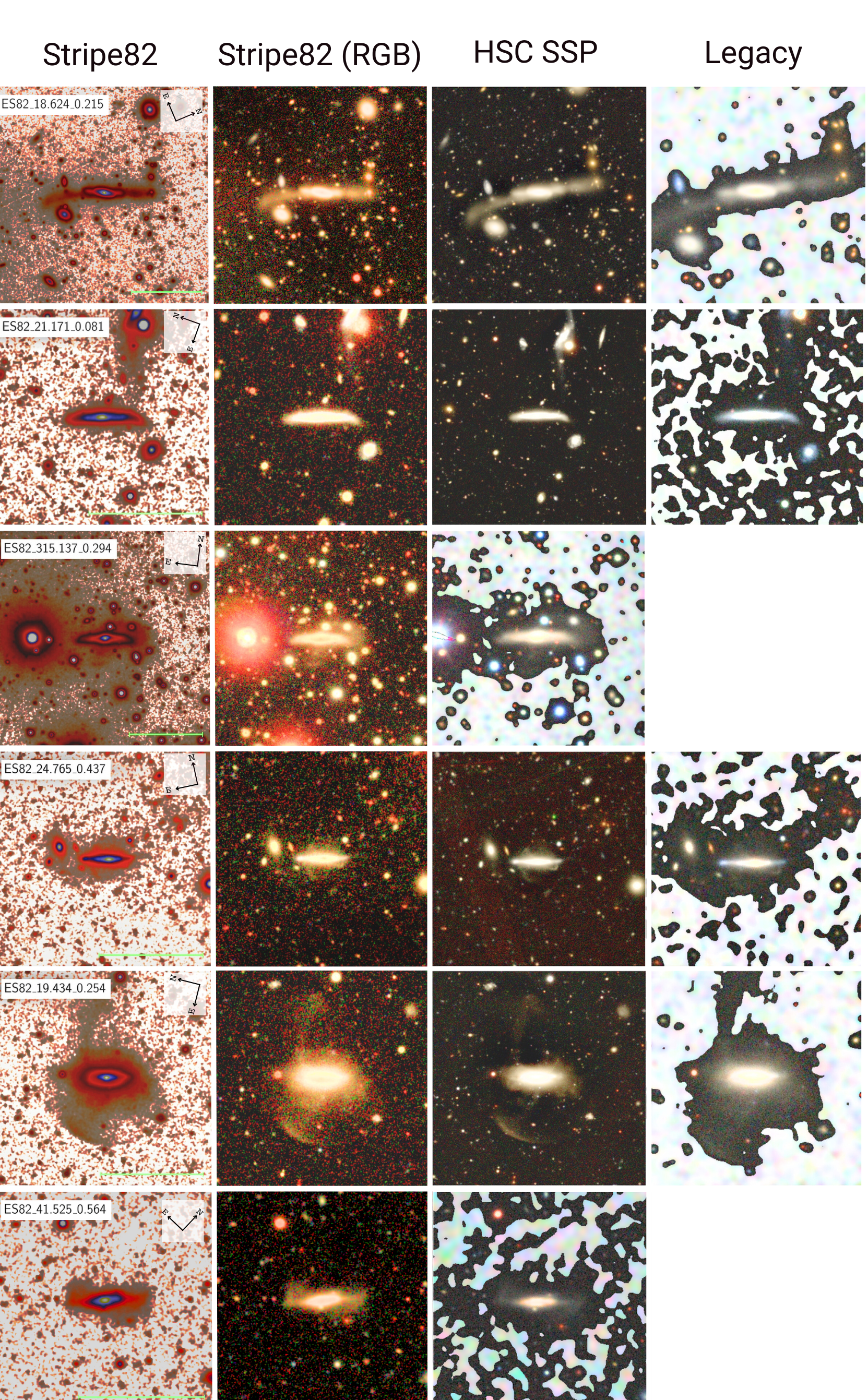}

\caption{Examples of observed tidal structures are demonstrated in three independent surveys: first two columns are Stripe 82 coadds and Stripe 82 RGB images (created using g,r,i filters), the third and fourth columns are images from HSC-SSP (coadds) and DESI Legacy surveys respectively. The first line demonstrates a galaxy with a tidal tail, the second line -- with a bridge, the third line -- with an arc structure, the fourth line -- with a satellite remnant, the fifth line -- with a shell, and the sixth line -- with the disc deformations. The scales of object images from different surveys may vary.}
\label{fig:tidal}
\end{figure*}

Fig.~\ref{fig:tidal} displays typical examples of galaxies with tidal structures from our sample. In this section, we present a concise overview of each galaxy from Fig.~\ref{fig:tidal}, offering valuable details about its structural characteristics derived from our personal examination of images from the three different deep surveys:\\

\textbf{ES82\_18.624\_0.215}

Also known as SDSS J011429.85+001254.7 (SDSS spectral redshift $z = 0.046$), according to \citep{2017A&A...599A..81M} this edge-on galaxy is contained in a compact group with the mean redshift $z = 0.045$. In Fig. ~\ref{fig:tidal}, one can see a tidal tail or streams in northwestern and southeastern directions. It can be reasoned that this galaxy had a tidal interaction with neighbouring object SDSS J011429.79+001216.0 with SDSS spectral redshift $z = 0.045$ and other galaxies from the cluster. In turn, other neighbouring galaxies do not show obvious signs of interaction with the object under study.\\

\textbf{ES82\_21.171\_0.081}

This edge-on galaxy ($z = 0.056$) is also known as SDSS~J012440.99+000452.1. According to HyperLeda \citep{2014A&A...570A..13M}, it has the SBd morphological type. \cite{2006A&A...445..765K} determined the object morphological class as Sd.  
In all four deep images, we clearly see a bridge that connects the object with its neighbour SDSS~J012437.41+000513.9 which is confirmed by its neighbour's SDSS spectral redshift $z = 0.056$. The bridge connects the north-west outskirts of ES82\_21.171\_0.081 with  SDSS~J012437.41+000513.9 to the east. Both objects have warped discs which also points to their tidal interaction. \\

\textbf{ES82\_315.137\_0.294}

ES82\_315.137\_0.294 or SDSS~J210032.89+001739.9 has the SBb morphological type and redshift $z = 0.050$. This object demonstrates an arc-shaped structure on the west side of the galaxy. The structure is seen in three images (HSC image is not available). The arc extends to the north and after that bends sharply in the opposite direction. It is noteworthy that there are no objects in the vicinity of the galaxy for tidal interaction (\citealt{2010ApJS..186..427N}). This arc can be a dwarf satellite stretched by the tidal forces of the host galaxy.\\

\textbf{ES82\_24.765\_0.437}

Another name for this object is SDSS~J013903.59+002611.2. The galaxy has an SBc HyperLeda morphological type or Scd type determined by \citet{2006A&A...445..765K} and redshift $z = 0.100$. The galaxy demonstrates a tidally disrupted low-mass satellite orbiting the body of the host galaxy that can finally be ingested by it. According to \citet{2007ApJ...671..153Y}, the object is located in a group. The neighbouring galaxy in a northeast direction does not have a spectroscopic redshift, but its photometric redshift is far greater (0.157) than that of the target galaxy.

\textbf{ES82\_19.434\_0.254}

Also known as SDSS~J011744.15+001515.5, has a redshift of $z=0.076$. It does not have a detailed morphological classification in HyperLeda, but its geometry and colour suggest that this is an S0 galaxy. A distinctive feature of this object is its pronounced shell that is a clear sign of a minor merger, which is when a small galaxy is swallowed by a larger one. There are no clearly seen galaxies in the vicinity to claim that the galaxy is located within a compact group. 

\textbf{ES82\_41.525\_0.564}

This edge-on galaxy, also known as 2MASX J02460608+0033487, exhibits distinct features of tidally distorted spirals and/or disc warp. This peculiar morphology suggests a potential interaction scenario with the neighbouring galaxy 2MASX J02460531+0036297, situated within a close proximity of 3 arcminutes from ES82\_41.525\_0.564. Notably, both galaxies share the same redshift value of $z = 0.075$ as determined from SDSS spectroscopic analysis. This concordance in redshift values further supports the hypothesis of a physical association or ongoing interaction between these two galactic systems.

\par
\par
In the second category of distinctive structural features, we include such structures as:
\begin{enumerate}
    \item Lopsided Discs: significant disc asymmetries along the plane relative to the centre (\citealt{1980MNRAS.193..313B}, \citealt{2005A&A...438..507B}, \citealt{2008A&ARv..15..189S}). 
    \item Thin Warps: a slight departure of the disc matter from the mean galactic plane which becomes especially prominent at the galaxy periphery (\citealt{1990MNRAS.246..458S}, \citealt{1998A&A...337....9R}, \citealt{2002A&A...382..513R}, \citealt{2017MNRAS.465.3446G}). The disc warps are typically much more pronounced in H{\sc I}, outside of the optical disc, but optical warps can be quite prominent as well \citep{1979A&AS...38...15V,2016MNRAS.461.4233R}.  
    A disc warp may arise from a misalignment between the disc and its surrounding dark halo, which can vary over time due to precession  (\citealt{2006ApJ...641L..33W}. Alternatively, the warp could result from interactions with satellites (\citealt{2022ApJ...935...48Z}). We subjectively divide these two types by the prominence and thickness of the warp and classify them into tidal (thick) warps and thin warps which may be produced by other mechanisms.
    \item Polar Structures: large-scale outer rings or discs of gas, dust and stars, orbiting in the plane approximately perpendicular to the disc of the main or host galaxy (\citealt{1990AJ....100.1489W}, \citealt{1997A&A...325..933R}, \citealt{2011MNRAS.418..244M}, \citealt{2015MNRAS.447.2287R}).
\end{enumerate}

\begin{figure*}
\centering
\includegraphics[width=15cm]{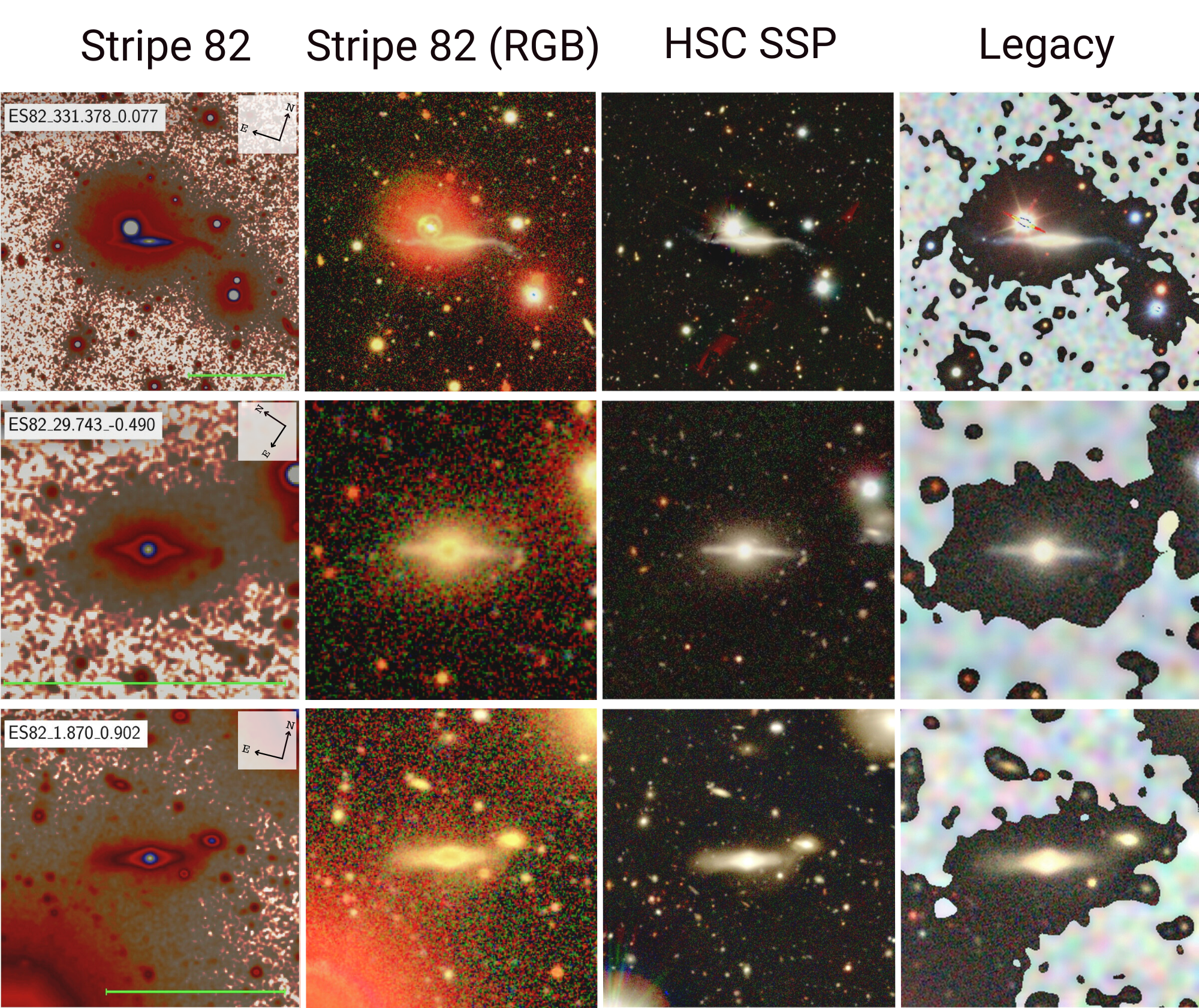}

\caption{Examples of observed structural features are demonstrated in three independent surveys: first two columns are Stripe\,82 coadds and Stripe\,82 RGB images, the third and fourth columns are images from HSC-SSP (coadds) and DESI Legacy surveys respectively. The upper row displays a galaxy with a warp, the middle row --- a candidate for polar-ring galaxy (PRG), and the bottom row --- a galaxy with a lopsided distorted disc.}
\label{fig:features}
\end{figure*}

Examples of structural features that we classified in our catalogue are illustrated in Fig.~\ref{fig:features}.

\textbf{ES82\_331.378\_0.077}

In this galaxy, the main prominent feature is the thin disc warp. Galactic warps may often be a consequence of galaxy-galaxy tidal interactions \citep{2002A&A...386..169L}. In our deep images, there are no close objects that could gravitationally interact with this object. \citet{2007A&A...470..505V} included this galaxy in their catalogue of isolated objects, so the warp was probably generated by some other mechanism. For example, dark matter halos can play a role in the formation of disc warps (\citealt{1988MNRAS.234..873S}, \citealt{2014A&A...572A..23C}, \citealt{2023ApJ...957L..24H}, \citealt{2023NatAs...7.1481H}).\\

\textbf{ES82\_29.743\_-0.490}

Another example of a galaxy with a prominent structural feature is the polar ring galaxy candidate ES82\_29.743\_-0.490 which is also called SPRC-77. \citet {2011MNRAS.418..244M} included this object in their catalogue of PRGs as a good candidate based on the galaxy's appearance. There are different mechanisms for PRG formation, such as major mergers, minor mergers, or gas accretion. On our deep images, this galaxy has a distinctive oval central component and a blue ring. This PRG also demonstrates a small south-eastern arc that can be satellite debris stretched during tidal interaction \citep{2022RAA....22k5003M}. 

\textbf{ES82\_1.870\_0.902}

Another structural feature that is demonstrated in Fig.~\ref{fig:features} is disc lopsidedness. We can clearly see that the disc in ES82\_1.870\_0.902 is severely disrupted: the galaxy's centre of symmetry is shifted in the northwest direction. It is believed that such non-asymmetry may originate from galaxy mergers \citep{1997ApJ...477..118Z}. We suppose that the lopsidedness of this galaxy is caused by the tidal interactions with its neighbour, PGC~170824, located north-west of the studied object (their redshifts are 0.100 and 0.102, respectively). It is not clear if there is a stellar bridge between these galaxies. Additionally, the centre of symmetry of ES82\_1.870\_0.902 is shifted towards PGC~170824. \\

In total, tidal structures and distinctive structural features are observed in 49 and 56 galaxies out of 838, respectively (taking into account Galactic cirrus, see below). Quantitative statistics by tidal structures are as follows: 
\begin{enumerate}
    \item Tidal tails and streams (15)
    \item Diffuse shells (6)
    \item Bridges (9)
    \item Arcs (7)
    \item Satellite debris (7)
    \item Disc deformations (8)
\end{enumerate}

In general, types of tidal structures are not limited to those listed above. These LSB structures are the most frequently encountered in our study (we refer the interested reader to, for example, \citealt{2010AJ....140..962M}, \citealt{2019A&A...632A.122M}, \citealt{2022A&A...662A.124S}). The parameters of galaxies with tidal structures are given in Table~\ref{table:table1}.

The numbers of galaxies with different structural features are as follows:
\begin{enumerate}
    \item Lopsided discs (11)
    \item Warps (45)
    \item Polar rings (3)
\end{enumerate}

As discussed briefly at the beginning, if a galaxy is contained within the Milky Way's cirrus clouds or filaments, then it can appear as if it has an LSB feature. To verify the results of our classification, we must ensure that the identified LSB features are not corrupted by Galactic cirrus. \citet{2023MNRAS.519.4735S} used SDSS Stripe 82 images to search for cirrus clouds with a surface brightness < 29 mag\,arcsec$^{-2}$ and compiled a cirrus map using a neural network that spans the entire Stripe 82. Fig.~\ref{fig:cirrus} in the Appendix shows where our sample of galaxies is located along the cirrus map. We determined that 72 galaxies in our whole catalogue are within 3 times their radius to a cirrus filament or are directly contained within one. Out of those 72 galaxies, 16 are directly contained within a cirrus filament. There are 8 galaxies with tidal features that have cirrus filaments within close proximity, and upon inspection, it was concluded that 7 of those galaxies had minimal contamination and contained legitimate tidal features. In contrast, the galaxy ES82\_2.265\_-0.583 (an apparent tail), was highly contaminated and its effect could not be ruled out and therefore had to be removed as a tidal feature candidate. Therefore, cirrus has no appreciable effect on the results of our classifications.       

\section{Photometric Decomposition}
\label{sec:decomp}

The axis scales of galaxies in our sample that had been determined using {\sc SExtractor} appeared to be incorrect for many of them. To redefine disc thickening, we carry out simple S\'ersic \citep{1968adga.book.....S} fitting to compare the values of apparent axis ratio for galaxies with and without LSB structures:

\begin{equation}\label{eq:1}
I(r) = I_\mathrm{e}\,\exp\Bigg[-k\bigg(\Big(\frac{r}{r_\mathrm{e}}\Big)^{\frac{1}{n}}-1\bigg)\Bigg]\,,
\end{equation}

\noindent where $I_\mathrm{e}$ denotes the surface brightness at the effective radius $r_\mathrm{e}$, the parameter $n$ is the S\'ersic index, and $k$ is a function of $n$ that ensures half of the total flux is enclosed within $r_\mathrm{e}$.

The decomposition is performed using the Python wrapper {\sc IMAN}\footnote{\url{https://bitbucket.org/mosenkov/iman_new/src/master/}} which utilises the {\sc GALFIT} \citep{2002AJ....124..266P} code. In turn, GALFIT  uses a Levenberg-Marquardt algorithm for $\chi^{2}$ minimisation. Utilising pre-existing images of galaxies in the $r$-deep filter (as detailed in Sect.~\ref{sec:data}), alongside masks and PSF images, we employ these inputs as data for our analysis. Consequently, we successfully retrieve essential model parameters for all but 3 galaxies within our sample, including the apparent axis ratios $q$, effective surface brightness levels, effective radii, S\'ersic index, integrated magnitude, and the diskyness/boxyness parameter $C_0$. When the value of $C_0$ is less than zero, the isophotes exhibit a disc-like appearance. On the other hand, they appear boxy when $C_0$ is greater than zero. If the isophotes can be accurately represented by pure ellipses, then $C_0$ equals zero. In cases where it was not possible to build an adequate model on the first try, we perform additional cropping or reduce the image region to fit, minimising the impact of masking inaccuracies. Three objects for which models have not been built (mostly because of starlight) are excluded from further analysis. The proportion of such galaxies is less than 1 per cent of the entire catalogue, so this aspect should not significantly affect the results. Additionally, we perform galaxy fitting with tidal structures masked to ensure that the interpretation of disc thickening does not strictly arise from the systematics in measuring $q$ for galaxies with tidal features. This procedure helps us answer the question, whether the light of tidal features affects the measurements of the axis ratio.

Since we are interested in an investigation of outer isophotes of sample galaxies with structures and without them, which are described by apparent axis ratio and $C_0$ parameter, we will dwell on them in more detail below. 

\subsection{The disc thickening}

\begin{figure*}
  \subcaptionbox*{}[.45\linewidth]{%
    \includegraphics[width=\linewidth]{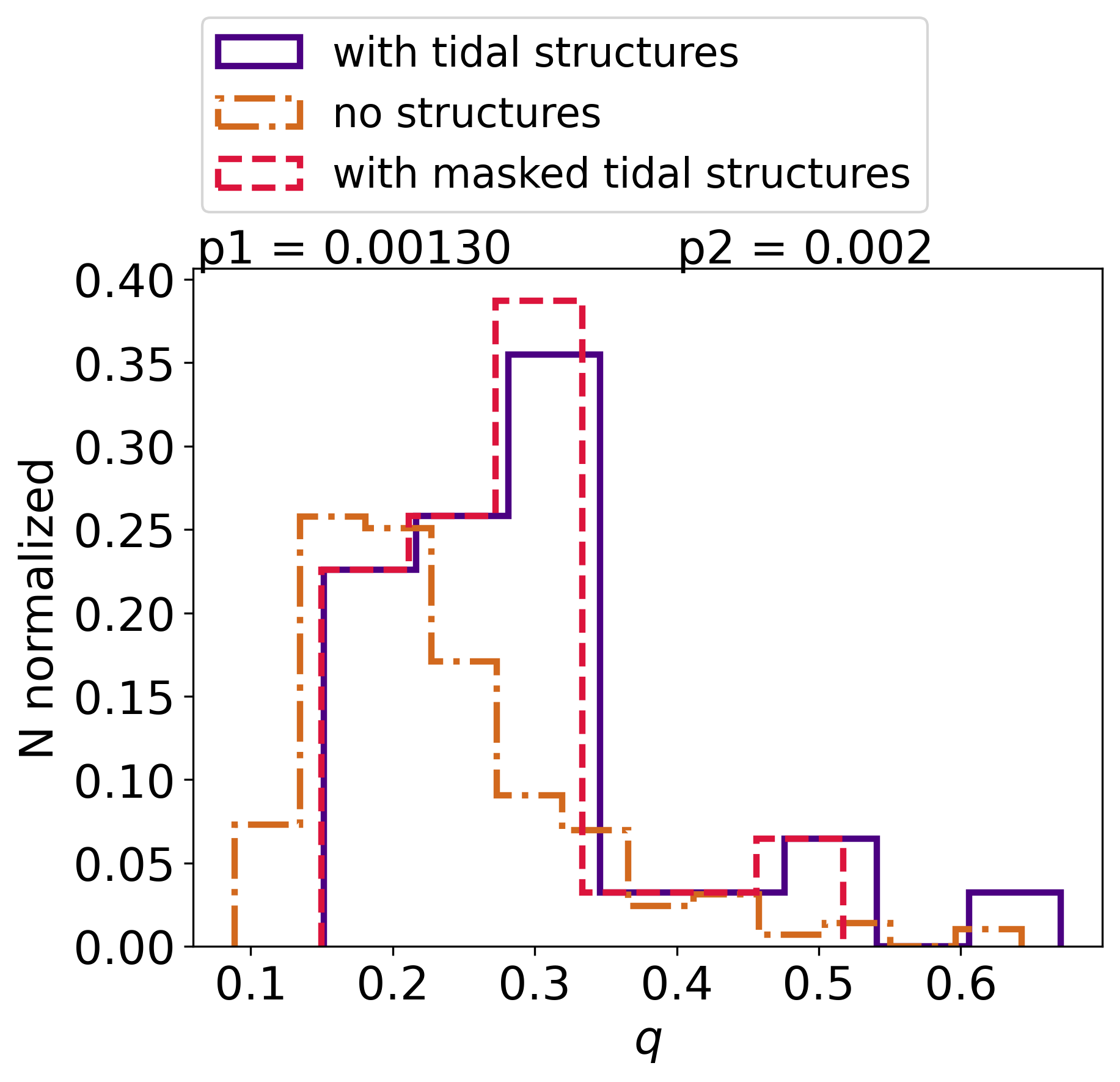}%
  }%
  \subcaptionbox*{}[.45\linewidth]{%
    \includegraphics[width=\linewidth]{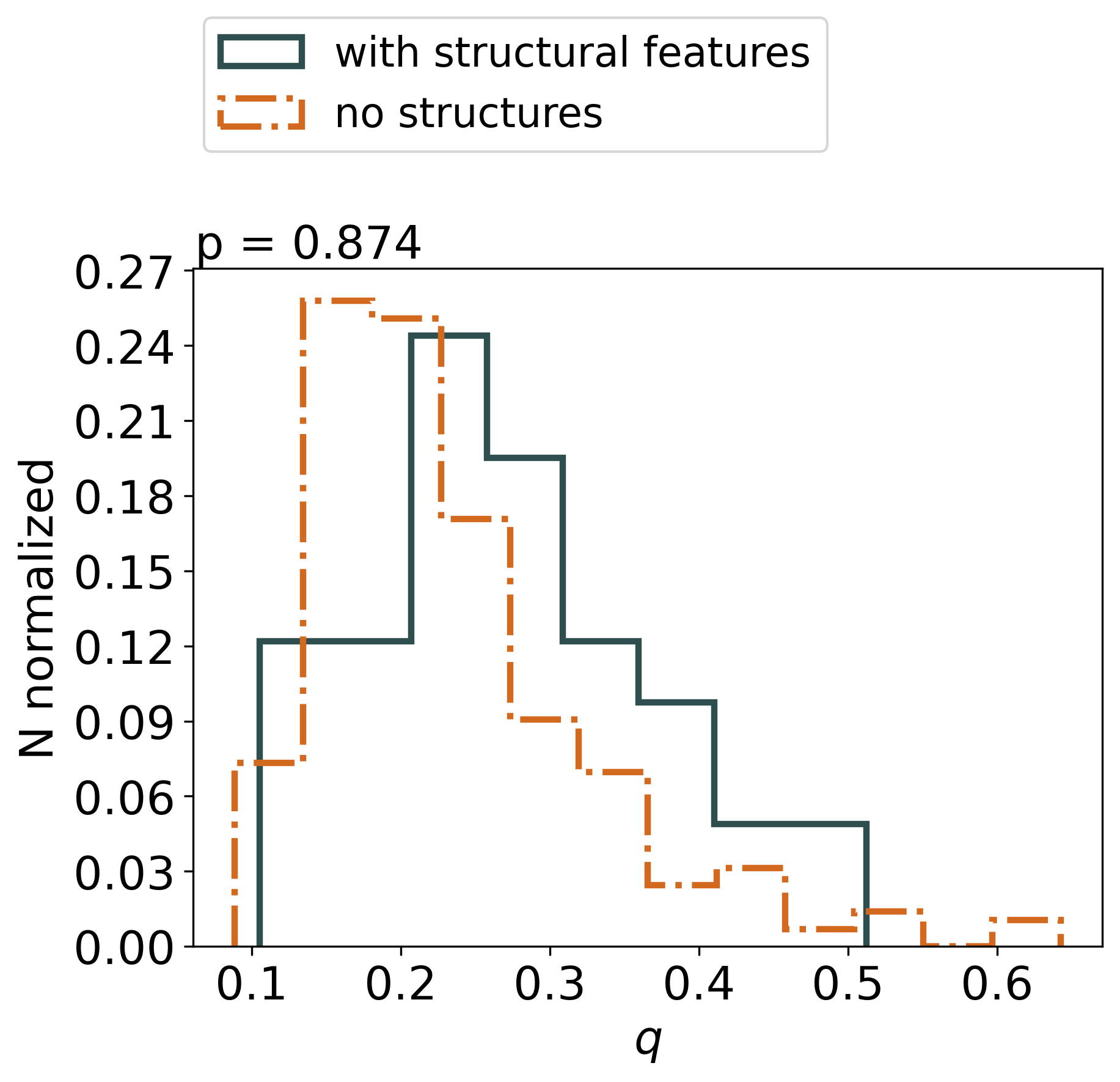}%
  }
  \caption{The left figure represents the axis ratio distribution of galaxies with tidal structures calculated using images with masked and unmasked structures and images of galaxies without any structural features. p1 is a p-value obtained from the Mann-Whitney test of the null hypothesis that the distribution underlying the sample of galaxies with tidal structures is the same as the distribution underlying the sample of galaxies without them. p2 is the p-value corresponding to the sample of galaxies with masked tidal structures and the sample of galaxies without structures. The right figure represents the axis ratio distribution of galaxies with structural features. p is a p-value obtained from the Mann-Whitney test of the null hypothesis that the distribution underlying the sample of galaxies with structural features is the same as the distribution underlying the sample of galaxies without them.}
  \label{fig:thickness}
\end{figure*}

In Tables~\ref{table:table1} and \ref{table:table2} in the Appendix, we list the model thickening $q$ for galaxies with observed tidal structures and structural features. To gain more reliable results, we divide catalogue objects into four groups, which are described in detail at the end of Sect.~\ref{sec:data}. In this section, we have excluded group 3 from the analysis to take into account the inclination effects. As shown in Fig.~\ref{fig:thickness}, the distributions of galaxies based on their axis ratios are not normal, therefore lower and upper quartiles are used to compare the disc thicknesses for subsamples combining the different groups. Table~\ref{table:q} presents the results of the analysis: 
 
    \begin{table*}
        \caption{The statistics for galaxy apparent axis ratio of subsamples with tidal structures (with and without structure masking), structural features, and without any visible structures for objects from different groups and their combinations, where $q$ is the disc axis ratio} obtained in the decomposition process.
    \renewcommand{\arraystretch}{1.5}
    \begin{tabular}{p{14mm}p{15mm}p{15mm}p{15mm}p{15mm}p{15mm}p{15mm}p{15mm}p{15mm}}
    \hline\hline 
          & \multicolumn{2}{c}{Tidal structures} & \multicolumn{2}{c}{Tidal masked} & \multicolumn{2}{c}{All structures} & \multicolumn{2}{c}{No structures} \\
    Group & $q$ &  Number & $q$ &  Number & $q$ & Number & $q$ & Number \\
    \hline
    1  &$0.17_{-0.01}^{+0.05}$ & 5 & $0.17_{-0.01}^{+0.05}$ & 5 & $0.2_{-0.05}^{+0.05}$ & 25 &  $0.19_{-0.03}^{+0.05}$ & 216 \\
    \hline
    2  & $0.29_{-0.03}^{+0.06}$ & 14 & $0.28_{-0.03}^{+0.06}$ & 14 & $0.28_{-0.05}^{+0.07}$ & 25 & $0.25_{-0.05}^{+0.05}$ & 382\\
    \hline
    4  & $0.29_{-0.04}^{+0.02}$ & 26 & $0.28_{-0.04}^{+0.02}$ & 26 & $0.3_{-0.04}^{+0.07}$& 44 & $0.28_{-0.05}^{+0.1}$ & 71\\
    \hline
    1 + 4  & $0.28_{-0.06}^{+0.02}$ & 31 & $0.28_{-0.06}^{+0.03}$ & 31 & $0.28_{-0.07}^{+0.03}$ & 69 & $0.21_{-0.04}^{+0.06}$ & 287\\
    \hline
    1 + 2 + 4 & $0.28_{-0.05}^{+0.03}$ & 45 & $0.28_{-0.06}^{+0.03}$ & 45 &  $0.28_{-0.06}^{+0.05}$ & 94 &  $0.23_{-0.05}^{+0.05}$ & 669\\

    \hline 
    \end{tabular}
        \label{table:q}
    \end{table*}
    
Regardless of which groups are used to analyse disc thickening, the subsample of galaxies without structures are still flatter than the subsample with tidal structures (see Table~\ref{table:q}). For a numerical assessment, we focus on the fourth row of the table, which presents the results based on galaxies from groups 1 and 4 as the inclusion of objects from the second group may influence the result due to the inclination effects. In groups individually, there may not be enough objects to obtain a reliable estimation. The median values in Table~\ref{table:q} show that galaxies with tidal structures and with all identified distinctive features (LSB and others) have about 1.33 times greater disc thickening than galaxies without them. 

To distinguish between the actual disc thickening and potential measurement errors when $q$ can be affected by present tidal feature light along the minor axis of the galaxy, we additionally utilise tidal structure masking during the fitting process. This allows us to quantify the level of enhancement in $q$ specifically attributable to genuine disc axis ratio. The results are presented in Fig.~\ref{fig:thickness} and Table~\ref{table:q}. Indeed, we can see in the distributions that there is some systematic effect of additional disc thickening due to tidal structures' light, but the real thickening of the galaxy disc is also present. In cases of fitting with and without structure masking, galaxies with tidal structures have about 1.33 times thicker discs than galaxies without them.

A control matching test is implemented to take into account the magnitude difference for the subsample of galaxies with tidal structures and without them. There is a probability that some galaxies without visible structures are fainter than galaxies with tidal LSB features. Therefore the first subsample may include galaxies whose discs are thickened by interactions, but whose features are too faint to be detected. To control for this possibility, a best-matching control is selected for each tidal feature object. The matching is performed using the values of the Kron radius (from the SExtracor catalogue) and model magnitudes in the $r$-band (from the SDSS database). The algorithm of the matching test is as follows:

Initially, the Kron radius and magnitudes are normalised. Subsequently, the square root of the sum of squared differences between corresponding parameters is computed for each galaxy pair. Ultimately, the total sum of these differences is evaluated:
$$Q = \frac{1}{N}\sum_{i=1}^N\frac{q_{ts}}{q_{ns}}$$ where $q_{ts}$ is a model thickness of galaxy with tidal structures, $q_{ns}$ is a matched galaxy without structures. $N$ is the number of galaxies with tidal structures in the sample. $Q$ shows how many times, on average, the disc thickening of galaxies with tidal structures differs from their best-matching controls. As a result of this test, we get that $Q = 1.46$ for model quantities obtained as a result of fitting without structure masking, and $Q = 1.39$ --- with structure masking. This test confirms the accuracy of the obtained results regarding disc thickening. Furthermore, the results from this test reveal even greater discrepancies in disc thickness among the subsamples under study.

Let us note that some galaxies in our sample exhibit an apparent axis ratio value $q$ greater than 0.3, which appears to contradict the criteria used for cataloging galaxies based on the SExtractor semi-major and semi-minor axes (denoted as the A\_IMAGE and B\_IMAGE parameters, respectively). This discrepancy can be attributed to the inaccuracies in the SExtractor measurements, which, in turn, affect the $q$ values used for the galaxy selection. While outliers that are clearly not edge-on galaxies have been manually filtered out, our sample still includes edge-on galaxies with $q$ values measured by GALFIT that exceed 0.3. This can be attributed to prominent galaxy bulges, bright halos, shells, thick stellar discs, and the contribution of the light from LSB structures or external objects.

In general, galaxies with structures do have thicker discs than galaxies without them. This difference is more significant when considering a subset of galaxies with only tidal structures. In our study, the difference in relative thickness is not as significant as in other works. For example, \citet{1997A&A...324...80R} reported that galaxies undergoing merging have a relative thickness that is, on average, twice that of non-interacting ones. Similarly, \citet{2000A&AS..144...85S} found this value to be approximately 1.7 times greater for interacting galaxies compared to non-interacting galaxies. However, these studies focused on real galaxy mergers in the midst of collision, whereas our research mostly involves galaxies that are either still too distant from each other to be considered colliding or are involved in minor mergers.

\subsection{The diskyness/boxyness parameter \texorpdfstring{$C_0$}{C0}}
\begin{figure*}
  \subcaptionbox*{}[.45\linewidth]{%
    \includegraphics[width=\linewidth]{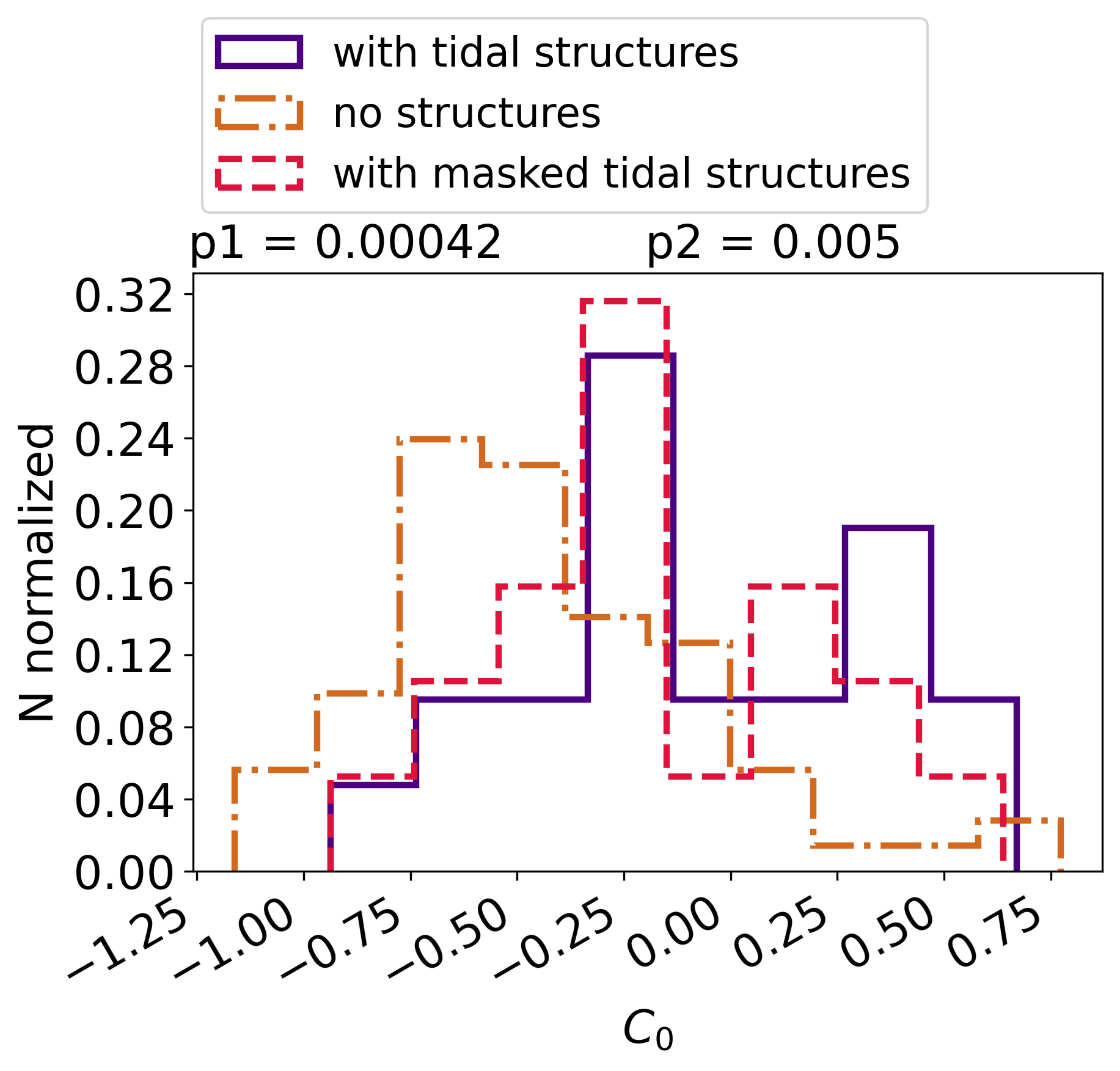}%
  }%
  \subcaptionbox*{}[.45\linewidth]{%
    \includegraphics[width=\linewidth]{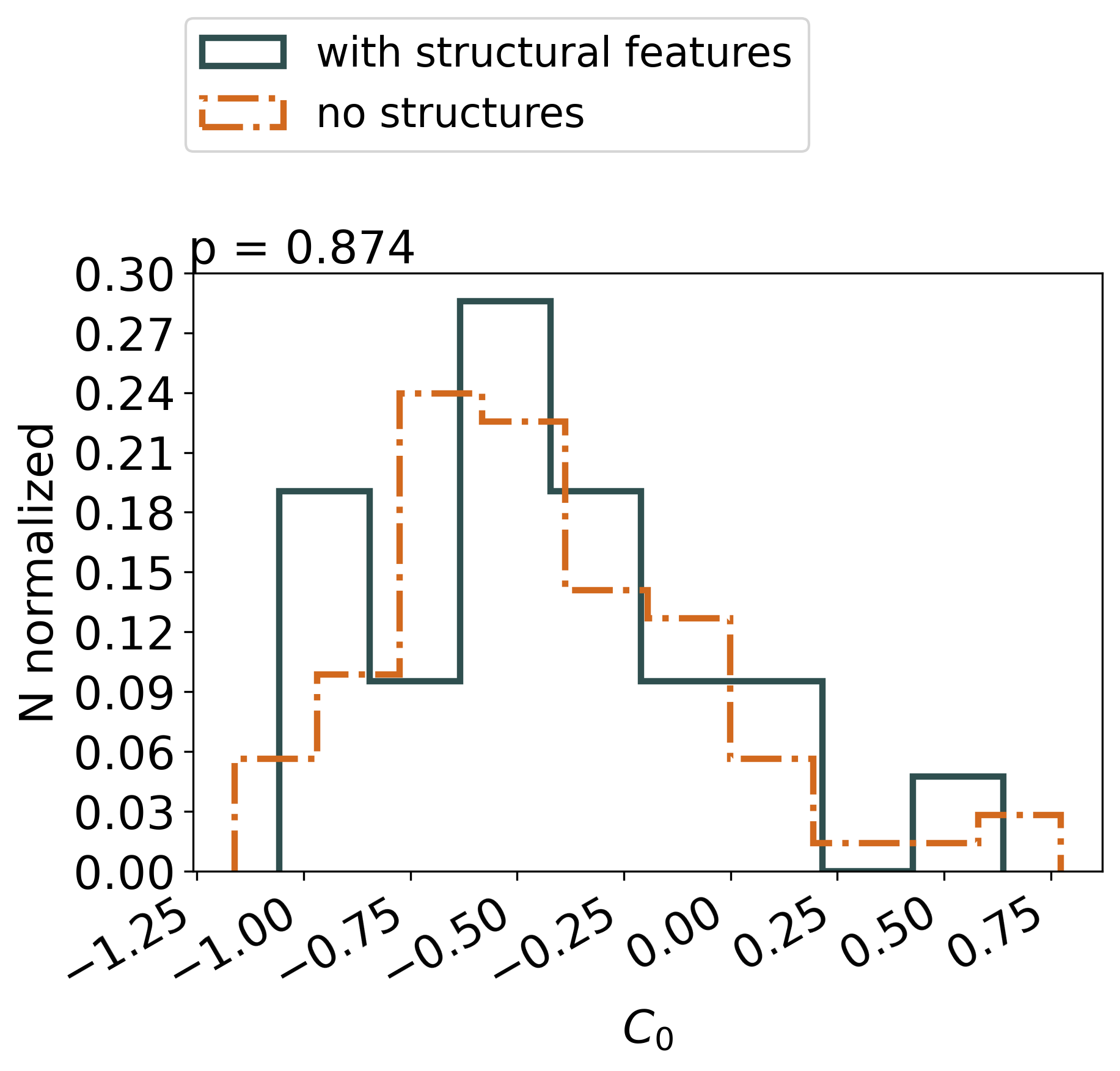}%
  }
  \caption{The left figure represents the $C_0$ distribution of galaxies with tidal structures calculated using images with masked and unmasked structures and images of galaxies without any structural features. p1 is a p-value obtained from the Mann-Whitney test of the null hypothesis that the distribution underlying the sample of galaxies with tidal structures is the same as the distribution underlying the sample of galaxies without them. p2 is the p-value corresponding to the sample of galaxies with masked tidal structures and the sample of galaxies without structures.} The right figure represents the $C_0$ of galaxies with structural features. p is a p-value obtained from the Mann-Whitney test of the null hypothesis that the distribution underlying the sample of galaxies with structural features is the same as the distribution underlying the sample of galaxies without them.
  \label{fig:c0}
\end{figure*}

The tidal interaction of galaxies with their environment can impact the shape of galactic outer isophotes, which in our investigation is characterised by the $C_0$ value. Galaxies with positive $C_0$ values demonstrate oval or boxy outer isophotes. In turn, galaxies with negative $C_0$ values have disky (diamond-like) external isophotes. In \citet{2020MNRAS.494.1751M}, it was shown that the mean value of $C_0$ for objects with LSB structures from the study sample is $\langle C_0 \rangle = 0.37 \pm 0.32$ versus $\langle C_0 \rangle = -0.29 \pm 0.40$ for the remaining ones. However, the sample from \citet{2020MNRAS.494.1751M} consists of only 35 galaxies, and the results of this paper should be taken with caution. We explored this issue in our catalogue that consists of 838 objects and can provide us with more reliable results. As in the part with the analysis of disc thickening, tidal structures are masked to avoid their influence on the resulting model. Fig.~\ref{fig:c0} shows the distribution of our galaxies by $C_0$. We can see that there is no significant difference in distributions for galaxies with structural features and galaxies without any structures. If we compare the diskyness/boxyness parameters for objects only with tidal structures and without them, we can see that the median of the distribution of galaxies with tidal structures is shifted towards positive $C_0$. In this case, the fraction of galaxies with tidal structures that have oval/boxy isophotes is larger which can also be seen in Fig.~\ref{fig:c0}. 

Table~\ref{table:C0} shows the statistics for parameter $C_0$ calculated for different groups and their combinations like it was done for apparent axis ratios. Here we also focus on the results described for the totality of groups 1 and 4. We can see from the table that for galaxies without any structures, the outer isophotes of the disc are indeed, on average, more oval/boxy than for galaxies with any kind of structures.

    \begin{table*}
    \caption{The statistics for galaxy $C_0$ value of subsamples with tidal structures (with and without structure masking), structural features, and without any observable structures for objects from different groups and their combinations. }
    \renewcommand{\arraystretch}{1.5}
    \begin{tabular}
    {p{14mm}p{15mm}p{15mm}p{15mm}p{15mm}p{15mm}p{15mm}p{15mm}p{15mm}}
    \hline\hline 
          & \multicolumn{2}{c}{Tidal structures} & \multicolumn{2}{c}{Tidal masked}  & \multicolumn{2}{c}{Structural features} & \multicolumn{2}{c}{No structures} \\
    Group & $C_0$ &  Number & $C_0$ &  Number &  $C_0$ & Number & $C_0$ & Number \\
    \hline
     1 & $-0.19_{-0.21}^{+0.3}$  & 5 & $-0.19_{-0.21}^{+0.3}$ & 5 & $-0.32_{-0.41}^{+0.2}$ & 25 &  $-0.49_{-0.2}^{+0.29}$  & 216 \\
    \hline
     2 & $-0.15_{-0.3}^{+0.16}$ & 14 & $-0.15_{-0.3}^{+0.16}$ & 14 &  $-0.3_{-0.22}^{+0.12}$ & 25 &  $-0.28_{-0.21}^{+0.35}$  & 382 \\
    \hline
     4 & $-0.15_{-0.1}^{+0.54}$ & 26 &  $-0.17_{-0.23}^{+0.26}$ & 26 & $-0.43_{-0.21}^{+0.21}$ & 44 & $-0.52_{-0.18}^{+0.21}$  & 71 \\
    \hline
     1 + 4 & $-0.15_{-0.19}^{+0.48}$ &  31 &$-0.18_{-0.23}^{+0.29}$ & 31  &  $-0.4_{-0.28}^{+0.27}$ & 69 &  $-0.5_{-0.2}^{+0.29}$  & 287 \\
    \hline
     1 + 2 + 4 & $-0.15_{-0.23}^{+0.28}$ & 45 & $-0.17_{-0.26}^{+0.24}$ & 45 &  $-0.38_{-0.29}^{+0.24}$ & 94 &  $-0.36_{-0.22}^{+0.31}$ & 669 \\
    \hline 
    \end{tabular}
        \label{table:C0}
    \end{table*}

\section{Discussion}
\label{sec:discussion}

\subsection{The Occurrence Rate of LSB Structures}

The main result of Sect.~\ref{sec:classification} with the classification and statistics of LSB and other distinctive features is that as compared to other works investigating galaxy formation models within the standard $\Lambda$CDM cosmological paradigm (\citealt{2001ApJ...557..137J}, \citealt{Martin2022}), we observe an insufficient number of galaxies with tidal structures in our catalogue. 
Interacting galaxies are a common phenomenon in the local Universe, as evidenced even by our own Galaxy (\citealt{1998Natur.394..752P}, \citealt{1999Natur.402...55L}, \citealt{2006ApJ...642L.137B}). Cosmological simulations within the standard $\Lambda$CDM model predict that a significant number of coherent tidal structures may be detected with sufficiently deep observations in the outskirts of the majority of nearby massive galaxies (\citealt{2011MNRAS.417.1260F}, \citealt{10.1093/mnras/stt1245}, \citealt{2015ApJ...799..184P}). Also, in \citet{2001ApJ...557..137J} where $\sim$100 parent galaxies are considered, it is shown (see their figure~8) that reaching a surface brightness limit $\sim$29 mag\,arcsec$^{-2}$ would reveal many tens of tidal features. Specifically, 20-40\% of galaxies from the sample should have approximately one detectable feature. In a recent study by \citet{Martin2022}, in preparation for the 10-year Legacy Survey of Space and Time (LSST) at the Vera C. Rubin Observatory, the ability to detect tidal features using the NEWHORIZONS cosmological simulation \citep{2021A&A...651A.109D} is investigated. As shown in figure~17 of \citet{Martin2022}, varying the limiting surface brightness in the $r$ band (28--31 mag arcsec$^{-2}$) affects the detection of tidal features across different redshifts. Using this figure and the mean redshift of our sample, we find that at a limiting surface brightness of 28 mag arcsec$^{-2}$ in the $r$ band, approximately 25\% of our sample should exhibit tidal features. Furthermore, \citet{2022arXiv220808443V} utilised the Magneticum simulation to measure a tidal feature fraction of 23\%, which is only 2\% lower than that found by \citet{Martin2022}. This comparison further demonstrates the insufficient number of galaxies with tidal features in our sample compared to those predicted by cosmological simulations.

We further seek to compare our results with those in the literature. It is important to note that none of the studies discussed below use the same method as this study to calculate the limiting photometric depth. Given the substantial differences in methodology, recalibrating the limiting surface brightness of our sample to align with those of the cited studies would require an extensive, time-intensive recalibration process. Consequently, such recalibration was deemed impractical and was not undertaken. In an observational study using HSC-SSP data, \citet{2018ApJ...866..103K} obtained the following result: tidal features were found in 1,201 out of 21,208 sample galaxies (5.6\%), which is consistent with our complete data set (galaxies with $R_{Kron}>61$~arcsec) confirming the presence of tidal structures in 26 galaxies out of 374 galaxies (7\%). However, their sample of galaxies at a surface brightness of $\mu_r = 28.1$~mag\,arcsec$^{-2}$ is incomplete. For our incomplete data set (galaxies with $R_{Kron} < 61$~arcsec), the presence of tidal features has a ratio of 24 galaxies out of 464 galaxies (5.2\%). \citet{2018ApJ...866..103K} used various methods obtained from the literature (see references therein) to calculate the limiting depth of their sample, but none of these methods match with our own. In \citet{2018A&A...614A.143M}, it is detected that tidal features are found around $\sim10$\% of a mass-selected sample of isolated Milky Way analogues at a limiting surface brightness of 28 mag\,arcsec$^{-2}$. Unlike the previous study, \citet{2018A&A...614A.143M} has a complete sample. To estimate the depth of their study, \citet{2018A&A...614A.143M} used 5$\sigma$ fluctuations with 3~arcscec diameter apertures. In a more recent study, \citet{2022A&A...662A.124S} found tidal features in 127 galaxies out of a sample of 352 galaxies (36\%, private communication), with features not extending beyond a photometric depth of 27.5\,mag\,arcsec$^{-2}$.


There could be numerous reasons why our statistics do not match with other observations and simulations. Reasons such as visual bias, projection effects, redshift and mass selection biases, insufficient image depth, contamination by Galactic cirrus clouds, and the reliability of cosmological hydrodynamic simulations can all play a role. Below, we briefly consider all of these.

As noted, visual classification was used to categorize the types of tidal and structural features within our sample. However, visual classification can be subjective, depending on the individual classifier, and can lead to disagreements among different classifiers (\citealt{2010ApJ...709.1067B}, \citealt{2020MNRAS.492.2075B}). For example, in \citet{Martin2022},five experienced classifiers participated in categorizing tidal features in mock images, where some classifiers marked an image as having a feature, while others marked it as featureless. When reviewing the images again, many revised their original answers. This shows that even classifiers with experience can face uncertainty in their classifications and that classifications should be revisited numerous times to ensure accuracy. In the case of our sample, it is likely that some galaxy images might be overlooked and not included as having tidal features. For this reason, the entire catalogue was examined by several classifiers to mitigate this subjective factor, but it is still difficult to estimate to what extent this has impacted our classification results. In \citet{2010ApJ...709.1067B}, a blind study found that classifications among different classifiers varied by only 3\%. These classifications were done on strong tidal signatures and, therefore, this percentage would be a lower estimate for our own study. Another method used to help eliminate visual bias is the method of applying weights to different classifiers and their classifications. Citizen science projects such as Galaxy Zoo and Galaxy Cruise use this method with great success, assigning greater weight to the most consistent classifications (\citealt{2017MNRAS.464.4420S}, \citealt{2023PASJ...75..986T}). In this study, we do not use this approach since we have only three well-trained classifiers.

In addition to that, by using multiple survey images in conjunction with those from the SDSS to identify tidal structures, galaxies were cross-referenced to ensure that structures are not image artefacts or any other non-extragalactic structures. This has not been extensively done in other studies and can serve as a way to curb visual bias. For example, the HSC-SSP, with its higher resolution images, helped us identify 10 more galaxies with tidal features. Unfortunately, only 64\% of the galaxies from our catalogue are covered by the HSC-SSP. Based on the additional galaxies identified among the 539 HSC-SSP galaxies, if we had HSC-SSP data for {\it all} catalogue galaxies, we could have found an additional $\sim6$ galaxies with tidal features, increasing our tidal fraction to about $7$\%.

Another factor that plays a role in visual classification bias is projection effects. A vast majority of studies done on tidal features use samples of galaxies with varying inclination angles, including the studies discussed above. The varying inclination angles can hide or change the shape of tidal features, making it difficult to provide accurate classifications \citep{Martin2022}. In \citet{Martin2022}, it was found that for deep imaging, the uncertainty from projection is dominant for their images. In the case of this paper, only edge-on galaxies are used, which helps to eliminate some projection effects and keeps the tidal classifications consistent. Due to this, projection effects do not play a significant role in our results.

As shown in Fig.~\ref{fig:z_spec} for galaxies with structures, the spectroscopic redshift distribution peaks at $z\simeq0.06$. Specifically, for tidal structures, the mean is $z = 0.07$ and it peaks at $z\approx0.04$. This is in agreement with observations and simulations (\citealt{2018ApJ...866..103K}, \citealt{Martin2022}, \citealt{
2023MNRAS.521.3861D}), which show that more tidal structures are identified at lower redshifts. This can be attributed to a combination of a loss of spatial resolution and cosmological dimming of the surface brightness at higher redshifts, which leads to greater difficulty in identifying LSB tidal structures \citep{2018ApJ...866..103K}. We performed a Mann-Whitney test to assess the null hypothesis that the distribution underlying the sample of redshifts from galaxies with structures is the same as that of the sample without structures. The p-value obtained from this test is $p = 0.07$. Therefore, we cannot reject the null hypothesis, concluding that the two sub-samples are similar. Analysing the redshifts from our sample using $(1+z)^3$ for AB magnitudes (see \citealt{2020ApJ...903...14W}) to account for cosmological dimming, the average is found to be 1.2 with a standard deviation of 0.15. This is not a significant enough change in the surface brightness; therefore, cosmological dimming should not play a significant role in our sample. Further evidence can be seen in Fig.~\ref{fig:surfbright} where the 
average offset between the measured tidal feature surface brightness and their intrinsic surface brightness before cosmological dimming is $\sim$ 0.09, which is not a significant offset. It is worth noting that the highest value for cosmological dimming occurs for a galaxy with a substructure. Looking at cosmological dimming values greater than or equal to 1.5, it is found that two galaxies with substructures met these criteria. Specifically, one of the two has a tidal structure and a cosmological dimming of $\sim2.3$ ($\Delta\mu = 0.36$), while the other has a dimming of 1.5. Therefore, cosmological dimming does not play a significant role in detecting tidal structures around our sample of galaxies and does not explain the lack of tidal structures in galaxy images.

\begin{figure}
\includegraphics[width=8cm]{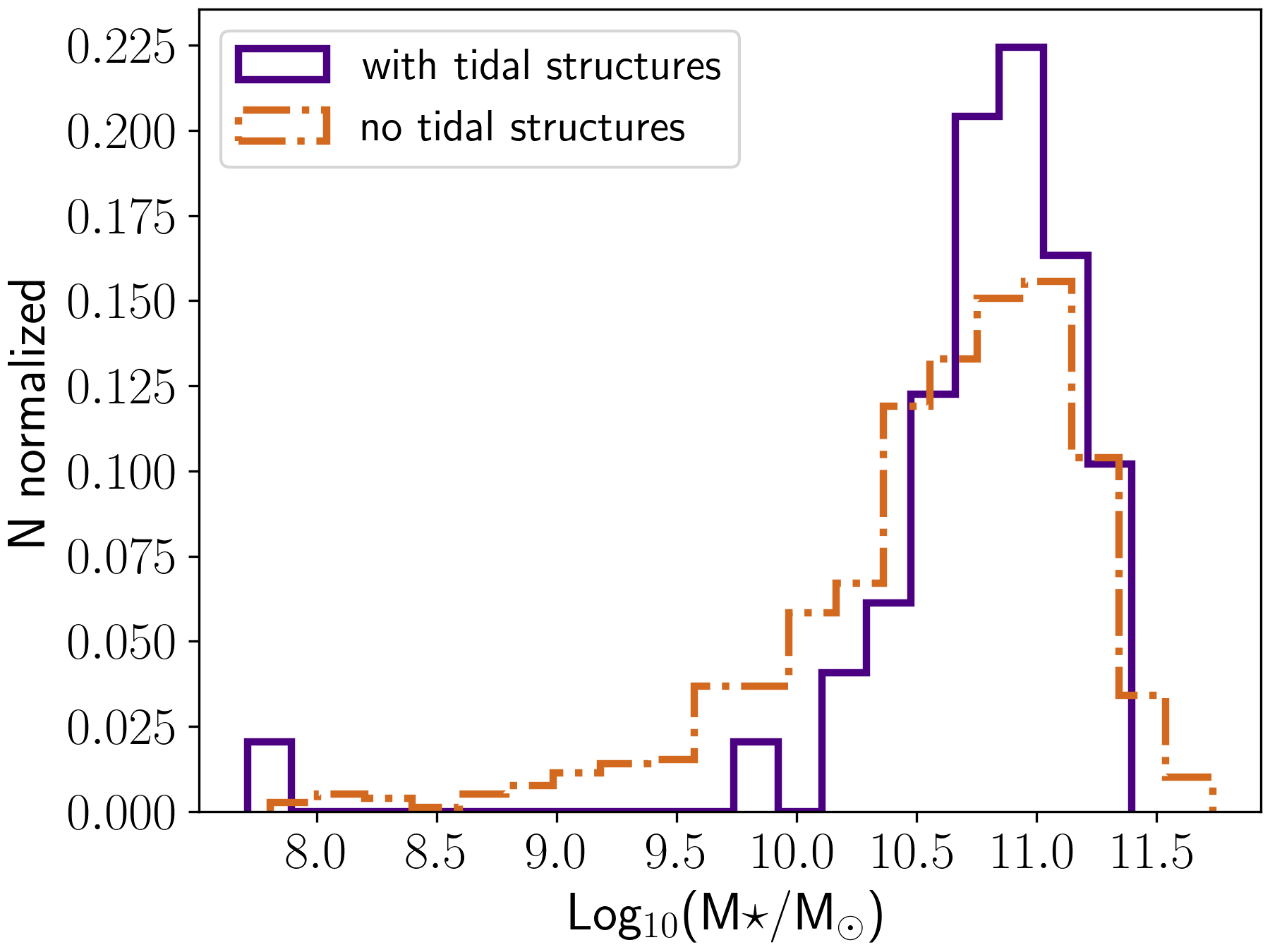}
\caption{Histogram comparing the masses of galaxies with tidal features and to the rest of our sample, including those with interesting features. Mass was calculated using the $g-r$ colour and the absolute magnitude in the $r$ band.}
\label{fig:mass}
\end{figure}

It has been found that the classification of tidal structures is mass-biased (\citealt{2013ApJ...765...28A}, \citealt{2018ApJ...866..103K}, \citealt{2020MNRAS.498.2138B}, \citealt{10.1093/mnras/stad131}). That is, more structures are found around larger mass galaxies. This is expected, as more massive galaxies tend to be more luminous as will the structures produced by minor/major interactions with these larger mass galaxies, making them easier to identify. Additionally, the number of major mergers that occur are expected to scale with the stellar mass of a galaxy, which means the number of substructures also scales with mass (\citealt{2023MNRAS.519.4920G}). The study by \citet{2020MNRAS.498.2138B} found that galaxies with stellar masses above $10^{11}\,M_\odot$ had 1.7 times more tidal structures than those below this value. Furthermore, in comparison with \citet{2020MNRAS.498.2138B}, \citet{10.1093/mnras/stad131} had 1.2 times more tidal structures. Our study suggests that the galaxies with more than one tidal feature have, on average, a stellar mass of $10^{10.6}\,M_\odot$. In Fig.~\ref{fig:mass}, galaxies with structures peak at $\sim10^{11}\,M_\odot$, in agreement with \citet{10.1093/mnras/stad131}. We performed a Mann-Whitney test of the null hypothesis that the distribution underlying the sample of galaxy masses with tidal structures is the same as the distribution underlying the sample of galaxy masses without tidal structures. The p-value obtained from this test is $p = 0.04$. Therefore, we can reject the null hypothesis and conclude that the two sub-samples are not similar. Therefore, there is a correlation between the stellar mass and the presence of tidal structures for our sample galaxies. The equation

\begin{equation}
\log_{10}\left(\frac{M}{L}\right) = a_{\lambda} + (b_{\lambda} \times \textrm{colour)}
\end{equation}

\noindent was used to find the stellar masses of our galaxies. The $a_{\lambda}$ and the $b_{\lambda}$ are -0.306 and 1.097, respectively, for the $r$ band and for the $g-r$ colour. The table of values and the equation above can be found in \citet{2003ApJS..149..289B}. It is important to note that the mass values calculated for our sample do not take into account dust attenuation in the galaxies but they are accurate enough for our general purposes.     

\begin{figure}
\includegraphics[width=8cm]{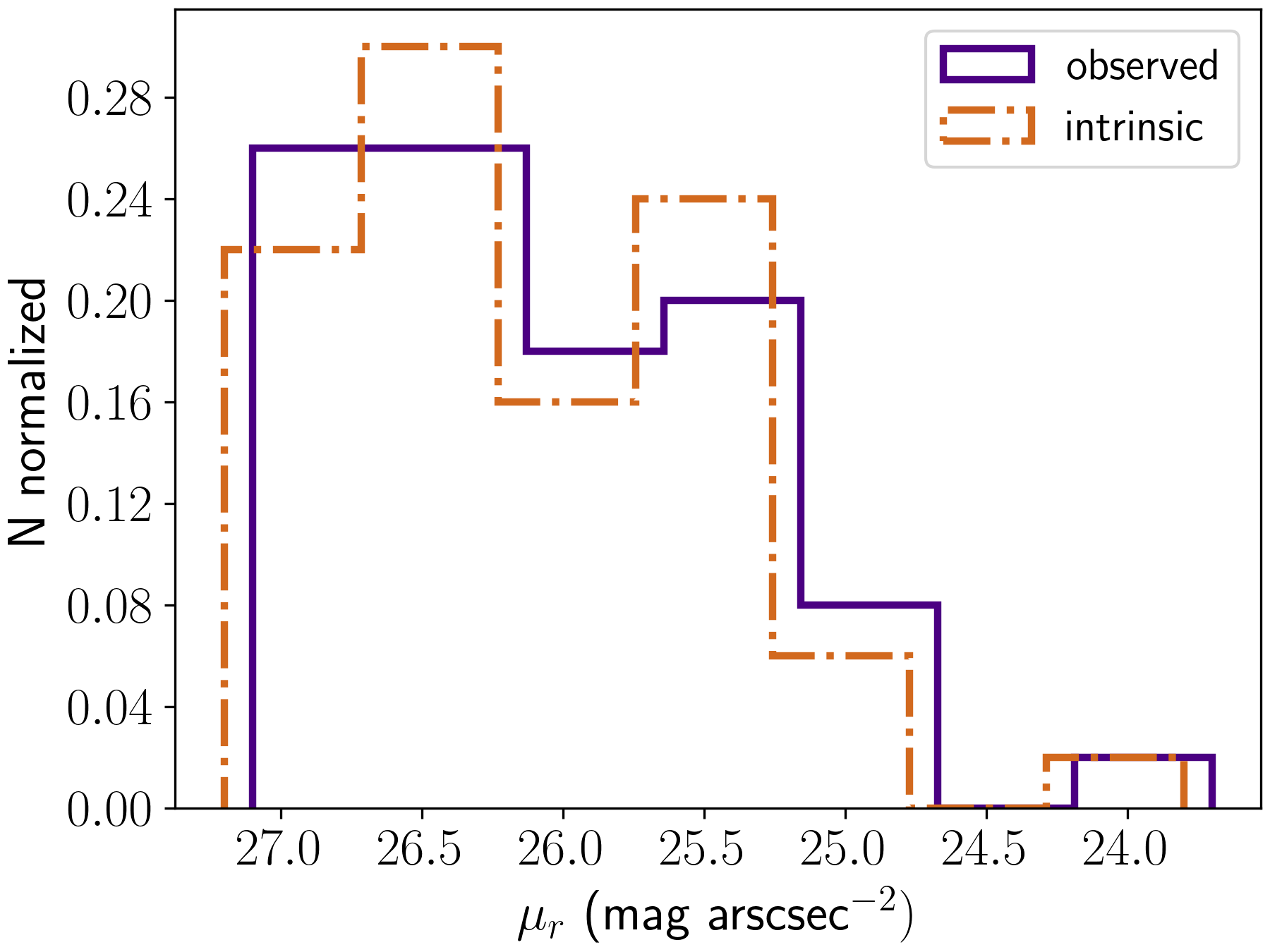}
\caption{Histogram of the median surface brightness of all tidal structures for the r-deep filter. This is overlaid with another histogram of the intrinsic surface brightness calculated using the cosmological dimming values for each galaxy. A small offset can be seen between the plots.}
\label{fig:surfbright}
\end{figure}
 
Another possible reason for the lack of tidal structures in the current study could be that our data is still not deep enough to detect the faintest LSB features. In \citet{2008ApJ...689..936J}, it is shown that the majority of tidal structures are expected to have a peak surface brightness higher than $\sim30$ mag\,arcsec$^{-2}$. However, as discussed above, \citet{2022A&A...662A.124S} found that the tidal features in their sample did not extend beyond the depth of a median value of 27.5\,mag\,arcsec$^{-2}$. We conducted a similar investigation into the depth of the tidal structures within our sample. By creating polygon regions around the structures themselves using SAO DS9, we were able to find that our sample of tidal structures does not extend beyond the photometric depth of a median value of 27.1 mag\,arcsec$^{-2}$ in the r-deep band (see Fig. ~\ref{fig:surfbright}). This result is similar to \citet{2022A&A...662A.124S} and could mean that these structures have not had enough time to accrete onto their host galaxy and dissipate, therefore remaining luminous. The mass of the galaxies may also play a role as all tidal structures in our sample have large masses and therefore are more luminous. This could also mean that our sample is still not deep enough, as only the more luminous structures are discernible, whereas the deeper structures are overlooked and missed completely. As shown in Fig.~\ref{fig:surfbright}, the measured surface brightness of the tidal structures has a peak at 26.5~mag\,arcsec$^{-2}$ and an average of 26.1~mag\,arcsec$^{-2}$ with a standard deviation of 0.7. In \citet{2022A&A...662A.124S}, their average is 25.6~mag\,arcsec$^{-2}$ and a standard deviation of 0.7 which matches our measurements.     

In summary, our catalogue of edge-on galaxies primarily contains high mass, low-redshift galaxies, examined in sufficiently deep images. Studies and cosmological simulations all agree that each of these parameters should produce the best chance of identifying a large fraction of tidal structures. Therefore, the reason for this study's low percentage could be a combination of some visual bias and our sample not being complete enough. Another reason not mentioned earlier could be that the predicted fraction of tidal structures that should be observed from cosmological simulations could be inaccurate. Even though cosmological simulations have had great success at producing observed characteristics of galaxies such as hierarchical structure formation and reproducing the general properties of galaxies (\citealt{Somerville2015}, \citealt{2022MNRAS.514.2936W}, \citealt{2024A&A...683A.182B}) they are not infallible and can produce discrepancies between prediction and observation. For example, the longstanding discrepancy of the star formation rate vs. stellar mass relation that gives a steeper slope than observed for lower mass galaxies at intermediate redshifts (\citealt{Daddi2007}, \citealt{Christensen2012}, \citealt{Somerville2015}). More recent discrepancies have been found such as the disparities in metallicity recycling and mixing history for larger galaxies \citep{2024MNRAS.530.1369J}. Discrepancies such as these possibly arise from not having a clear understanding of the mechanisms that govern specific physical processes such as star formation and stellar feedback (\citealt{Somerville2015}). This lack of understanding could explain why the few studies discussed above and our own do not match cosmological simulations. 

Additionally, simulations such as the one used in \citet{Martin2022} produce `smooth' images or images that do not contain contaminants, background/foreground objects (stars and galaxies), sky subtraction residuals, imaging artefacts, etc. \citep{2023MNRAS.521.3861D}. This allows the classifiers to know that the structures seen in the mock images are true structures, leading to higher fraction counts. Another point to consider is the morphology of the galaxies being produced in simulations. Disc galaxies or late-type galaxies (LTGs) have been found to be more easily disrupted and will produce tidal features more easily than compact spheroids (\citealt{2009ApJS..182..216K}, \citealt{2014A&A...567A..47P}). If a simulation produces more disc galaxies than spheroidal galaxies, then it is possible that the simulation would produce a higher fraction of tidal features. Looking specifically at the IllustrisTNG (\citealt{2018MNRAS.480.5113M}, \citealt{2018MNRAS.477.1206N}, \citealt{2018MNRAS.475..624N}, \citealt{2018MNRAS.475..648P}, \citealt{2018MNRAS.475..676S}, \citealt{2019ComAC...6....2N}) and EAGLE (\citealt{2015MNRAS.450.1937C}, \citealt{2015MNRAS.446..521S}) simulations, it was found that the morphology of the simulated galaxies matched observations, where only the asymmetry is larger for the simulated galaxies (\citealt{2020MNRAS.491.3624B}, \citealt{2023MNRAS.519.4920G}). The NEWHORIZON simulation used by \citet{Martin2022} also agrees well with the observed morphology \citep{2021A&A...651A.109D}. This means that the morphology in simulations does not account for the observed discrepancy.

This discussion prompted us to look at the morphology of the galaxies in our sample using the HyperLeda database. For the entire sample, 519 (62\%) are LTGs, 295 (35.2\%) are early-type galaxies (ETGs), and 24 (2.9\%) have an unknown morphology. Looking at only tidal structures, 32 galaxies (65.3\%) are LTGs, and 17 (21.1\%) are ETGs. Eleven of the LTGs had uncertain classifications and were further classified by eye. Of the three galaxies that contain more than one tidal structure, two are ETGs and one is a LTG. It is known that more tidal features are found around ETGs due to their formation and evolution mechanism being minor and major mergers which produce more easily visible features such as plumes \citep{2017IAUS..321..180D} and shells \citep{2020ApJ...905..154Y}. With the majority of our sample consisting of LTGs, another reason for the discrepancy could be that our sample lacks ETGs. Many of the studies mentioned focus either solely on ETGs or have a more balanced split between ETGs and LTGs. Additionally, these studies often have varying surface brightness limits. Both effects may lead to different tidal feature fractions.

It is difficult to pinpoint the exact reason for the discrepancy between the simulations and observations, but our discussion above suggests that further investigation is needed. Deep photometric studies of large, complete samples of galaxies should better describe the statistics of tidal structures in the local Universe for a more appropriate comparison with cosmological simulations. In one such future study, we will examine almost 6,000 true edge-on galaxies from the Catalogue of Edge-on Disc Galaxies from SDSS \citep{2014ApJ...787...24B}. This will be the largest catalogue of edge-on galaxies with tidal/structural features ever compiled. Further statistical analysis will be performed and compared with the results from this study.

\subsection{Disc thickening}

Galaxy mergers are believed to be the main drivers for significant kinematic perturbations of galaxies (\citealt{1972ApJ...178..623T}, \citealt{2015A&A...582A..21B}). This fact is confirmed in theoretical (e.g \citealt{1992ApJ...400..460H}) and observational (e.g., \citealt{2006AJ....132..197W}, \citealt{2007AJ....134..527W}) studies. Major mergers, in particular, can significantly change the morphology of a galaxy \citep{2012MNRAS.423.1544S}. However, such events are not as common as minor mergers or interactions of satellite galaxies, which can also impact the disc structure \citep{2008ApJ...688..254K}. 

One of the types of deformations caused by perturbations in the disc as a result of tidal interactions is the thickening of the disc. Many simulations show that these tidal interactions trigger instabilities in the disc and lead to its fattening (e.g., \citealt{1999ApJ...524L..19M}). Important results from galaxy-galaxy encounters, even those not leading to a merger, are obtained in N-body simulations by, for example, \citet{1990A&A...230...37G}. The authors demonstrate that interactions with a neighbour, orbiting perpendicular to the plane of the target galaxy, produce a bending and subsequent disc thickening. Such interactions lead to a greater effect than in-plane interactions. For the most part, early simulations predict disc thickening by a factor of 2--4 (\citealt{1990ASPC...10...25O}, \citealt{1992ApJ...389....5T}, \citealt{1993ApJ...403...74Q}, \citealt{1995ApJ...447L..87M}). At the same time,  in the work of \citet{1999MNRAS.304..254V}, it is shown that the impact of tidal heating is overestimated. The increase of the vertical scale height associated with this effect is greater by a factor of 1.5--2, which is closer to the results obtained in observations.   

In discussing the results obtained in observational studies, one can refer, for example, to the works of \citet{1996A&AS..116..417R} and 
\citet{1997A&A...324...80R}. These authors investigate the effects of tidally-triggered disc thickening between galaxies of comparable mass. Using optical photometric data for a sample of 24 interacting galaxies and a control sample of 7 non-interacting disc galaxies, they find that the radial-to-vertical scales ratio $h/z_0$ for the first sample is twice as small as that for the second one, which confirms the simulation results. In \citet{2000A&AS..144...85S}, the authors also observe differences in $h/z_0$ values for interacting and non-interacting galaxies. In this study, the apparent axis ratio value of galaxies for the first group is found to be $\approx1.7$ times smaller than that for the second group. 

In accordance with the results of our work described in Sect.~\ref{sec:decomp}, the discs of galaxies with observable low-surface brightness structures are 1.33 times thicker than those of galaxies without observable structures. This value is slightly less than those obtained in the aforementioned studies. This difference is due to the fact that those works focused on major interactions, while our sample consists mostly of minor interactions, which do not contribute as significantly to the thickening of the discs. Minor interactions leave specific tidal tracers and, in the context of this article, include streams/arcs, shells, bridges, and satellite debris (\citealt{duc14}, \citealt{2015MNRAS.454.2472H}). Such tidal structures are found around 86\% of the galaxies that exhibit tidal structures; therefore, it is safe to say that our sample of LSB structures consists mostly of minor interactions. 

\section{SUMMARY AND CONCLUSIONS}
\label{sec:summary}
In this paper, we created a catalogue of edge-on galaxies in deep SDSS Stripe\,82. Also, we identified and classified LSB and other distinctive structures near or in these galaxies based on data from three independent deep sky surveys: SDSS Stripe\,82, HSC-SSP DR3, and DESI Legacy Imaging Surveys DR9. 539 out of 838 galaxies in our catalogue have HSC-SSP images and 827 galaxies have data in DESI Legacy. This means that for at least 65\% of the galaxy sample, we have three different data sources for structure classification. Also, we investigated the thickening and boxyness/discyness of the galaxy discs depending on the structure's presence or absence. The conclusions of this study can be summarised as follows:

\begin{enumerate}

    \item A catalogue of 838 edge-on galaxies in the SDSS Stripe\,82 (\textit{ES82}) has been created.  
    The main characteristics of the galaxies have been obtained. The mean redshift of the galaxies we have considered is $\langle z_\mathrm{spec} \rangle$ = 0.06$^{+0.03}_{-0.02}$ (Fig. \ref{fig:z_spec}). The mean value of absolute magnitude for the sample galaxies is $\langle M_{r} \rangle$ = -20.7$^{+0.8}_{-0.7}$ (Fig. \ref{fig:M_abs}), which is consistent with the results of \citet{2014ApJ...787...24B} for the EGIS catalogue.
    
    \item The classification of galaxies by the presence of structures has been performed. It is important to note that the use of the three independent surveys played a significant role not only in classifying structures but also in determining the inclination of galaxies, especially due to the better angular resolution of HSC-SSP. The coadded and RGB images of SDSS Stripe\,82 were useful for selecting a primary sample of galaxies by their disc flattening and for marking those with unusual structures. Comparing images from SDSS Stripe\,82 with those from HSC-SSP and DESI Legacy allowed us to identify the presence of spiral structures in the discs and allocate such objects to a separate category, as galaxies close to an edge-on orientation affect the accuracy of the results obtained in Sect.~\ref{sec:decomp}. The use of the three surveys also increases the reliability of the fact that the observed features of galaxies are indeed real structures and not artefacts in the image.
    
    \item There are 49 (5.8\% of the sample) galaxies with LSB tidal structures (tails, bridges, arcs, shells, satellite remnants, disc deformations) and 56 (6.7\% of the sample) galaxies with distinctive structural features (thin disc warps, lopsided discs, and polar structures) are selected and classified. An important result of this work is that, despite the depth of the studied images, the number of galaxies with LSB structures in the sample is much lower than in other studies considering galaxy formation models that adopt the $\Lambda$CDM paradigm. A possible reason may be the mean size of galaxies in our catalogue. Our galaxies are mostly small, and possible faint small-size/thin structures can be difficult to distinguish in them. Another reason could be that our imaging is not deep enough and there could be visual bias. We also came to the conclusion that cirrus clouds and cosmological dimming do not play a significant role in our structure statistics.

    \item Simple S\'ersic fitting is performed for the entire sample. It is shown that galaxies with observed tidal structures are, on average, thicker (by 1.33 times) than the galaxies without these structures (Fig. \ref{fig:thickness}). Also, the discs of galaxies with tidal structures or structural features, on average, have more of a boxy-like shape of the outer isophotes than objects without any visible structures (Fig. \ref{fig:c0}).
\end{enumerate}

Based on the results and discussion of this paper, a large portion of studies done on LSB tidal features have small sample sizes, and larger studies are needed to get a better representation of the population of tidal features. In a future paper, we will continue to investigate the statistics, structure, and surface brightness of LSB tidal and structural features around edge-on galaxies using a large sample size of almost 6,000 edge-on galaxies. We hope this will provide new insights into these structures.

\section*{Acknowledgements}
We thank Elisabeth Sola for providing the statistics for their tidal structures from their study \citet{2022A&A...662A.124S}, which will be included in a future paper.

This research has made use of the NASA/IPAC Extragalactic Database (NED; \url{https://ned.ipac.caltech.edu/}) operated by the Jet Propulsion Laboratory, California Institute of Technology, under contract with the National Aeronautics and Space Administration. This research has made use of the HyperLEDA database (\url{http://leda.univ-lyon1.fr/}; \citealp{2014A&A...570A..13M}). 

Funding for the Sloan Digital Sky Survey IV has been provided by the Alfred P. Sloan Foundation, the U.S. Department of Energy Office of Science, and the Participating Institutions. SDSS-IV acknowledges
support and resources from the Centre for High-Performance Computing at the University of Utah. The SDSS website is www.sdss.org.

SDSS-IV is managed by the Astrophysical Research Consortium for the 
Participating Institutions of the SDSS Collaboration including the 
Brazilian Participation Group, the Carnegie Institution for Science, 
Carnegie Mellon University, the Chilean Participation Group, the French Participation Group, Harvard-Smithsonian Centre for Astrophysics, 
Instituto de Astrof\'isica de Canarias, The Johns Hopkins University, Kavli Institute for the Physics and Mathematics of the Universe (IPMU) / 
University of Tokyo, the Korean Participation Group, Lawrence Berkeley National Laboratory, 
Leibniz Institut f\"ur Astrophysik Potsdam (AIP),  
Max-Planck-Institut f\"ur Astronomie (MPIA Heidelberg), 
Max-Planck-Institut f\"ur Astrophysik (MPA Garching), 
Max-Planck-Institut f\"ur Extraterrestrische Physik (MPE), 
National Astronomical Observatories of China, New Mexico State University, 
New York University, University of Notre Dame, 
Observat\'ario Nacional / MCTI, The Ohio State University, 
Pennsylvania State University, Shanghai Astronomical Observatory, 
United Kingdom Participation Group,
Universidad Nacional Aut\'onoma de M\'exico, University of Arisona, 
University of Colorado Boulder, University of Oxford, University of Portsmouth, 
University of Utah, University of Virginia, University of Washington, University of Wisconsin, 
Vanderbilt University, and Yale University.

The Legacy Surveys consist of three individual and complementary projects: the Dark Energy Camera Legacy Survey (DECaLS; NOAO Proposal ID \# 2014B-0404; PIs: David Schlegel and Arjun Dey), the Beijing-Arisona Sky Survey (BASS; NOAO Proposal ID \# 2015A-0801; PIs: Zhou Xu and Xiaohui Fan), and the Mayall z-band Legacy Survey (MzLS; NOAO Proposal ID \# 2016A-0453; PI: Arjun Dey). DECaLS, BASS and MzLS together include data obtained, respectively, at the Blanco telescope, Cerro Tololo Inter-American Observatory, National Optical Astronomy Observatory (NOAO); the Bok telescope, Steward Observatory, University of Arisona; and the Mayall telescope, Kitt Peak National Observatory, NOAO. The Legacy Surveys project is honored to be permitted to conduct astronomical research on Iolkam Du'ag (Kitt Peak), a mountain with particular significance to the Tohono O'odham Nation.

NOAO is operated by the Association of Universities for Research in Astronomy (AURA) under a cooperative agreement with the National Science Foundation.

This project used data obtained with the Dark Energy Camera (DECam), which was constructed by the Dark Energy Survey (DES) collaboration. Funding for the DES Projects has been provided by the U.S. Department of Energy, the U.S. National Science Foundation, the Ministry of Science and Education of Spain, the Science and Technology Facilities Council of the United Kingdom, the Higher Education Funding Council for England, the National Centre for Supercomputing Applications at the University of Illinois at Urbana-Champaign, the Kavli Institute of Cosmological Physics at the University of Chicago, Centre for Cosmology and Astro-Particle Physics at the Ohio State University, the Mitchell Institute for Fundamental Physics and Astronomy at Texas A\&M University, Financiadora de Estudos e Projetos, Fundacao Carlos Chagas Filho de Amparo, Financiadora de Estudos e Projetos, Fundacao Carlos Chagas Filho de Amparo a Pesquisa do Estado do Rio de Janeiro, Conselho Nacional de Desenvolvimento Cientifico e Tecnologico and the Ministerio da Ciencia, Tecnologia e Inovacao, the Deutsche Forschungsgemeinschaft and the Collaborating Institutions in the Dark Energy Survey. The Collaborating Institutions are Argonne National Laboratory, the University of California at Santa Cruz, the University of Cambridge, Centro de Investigaciones Energeticas, Medioambientales y Tecnologicas-Madrid, the University of Chicago, University College London, the DES-Brazil Consortium, the University of Edinburgh, the Eidgenossische Technische Hochschule (ETH) Zurich, Fermi National Accelerator Laboratory, the University of Illinois at Urbana-Champaign, the Institut de Ciencies de l'Espai (IEEC/CSIC), the Institut de Fisica d'Altes Energies, Lawrence Berkeley National Laboratory, the Ludwig-Maximilians Universitat Munchen and the associated Excellence Cluster Universe, the University of Michigan, the National Optical Astronomy Observatory, the University of Nottingham, the Ohio State University, the University of Pennsylvania, the University of Portsmouth, SLAC National Accelerator Laboratory, Stanford University, the University of Sussex, and Texas A\&M University.

BASS is a key project of the Telescope Access Program (TAP), which has been funded by the National Astronomical Observatories of China, the Chinese Academy of Sciences (the Strategic Priority Research Program "The Emergence of Cosmological Structures" Grant \# XDB09000000), and the Special Fund for Astronomy from the Ministry of Finance. The BASS is also supported by the External Cooperation Program of Chinese Academy of Sciences (Grant \# 114A11KYSB20160057), and Chinese National Natural Science Foundation (Grant \# 11433005).

The Legacy Survey team makes use of data products from the Near-Earth Object Wide-field Infrared Survey Explorer (NEOWISE), which is a project of the Jet Propulsion Laboratory/California Institute of Technology. NEOWISE is funded by the National Aeronautics and Space Administration.

The Legacy Surveys imaging of the DESI footprint is supported by the Director, Office of Science, Office of High Energy Physics of the U.S. Department of Energy under Contract No. DE-AC02-05CH1123, by the National Energy Research Scientific Computing Centre, a DOE Office of Science User Facility under the same contract; and by the U.S. National Science Foundation, Division of Astronomical Sciences under Contract No. AST-0950945 to NOAO.

The Hyper Suprime-Cam (HSC) collaboration includes the astronomical communities of Japan and Taiwan, and Princeton University. The HSC instrumentation and software were developed by the National Astronomical Observatory of Japan (NAOJ), the Kavli Institute for the Physics and Mathematics of the Universe (Kavli IPMU), the University of Tokyo, the High Energy Accelerator Research Organisation (KEK), the Academia Sinica Institute for Astronomy and Astrophysics in Taiwan (ASIAA), and Princeton University. Funding was contributed by the FIRST program from the Japanese Cabinet Office, the Ministry of Education, Culture, Sports, Science and Technology (MEXT), the Japan Society for the Promotion of Science (JSPS), Japan Science and Technology Agency (JST), the Toray Science Foundation, NAOJ, Kavli IPMU, KEK, ASIAA, and Princeton University. 

This paper makes use of software developed for Vera C. Rubin Observatory. We thank the Rubin Observatory for making their code available as free software at http://pipelines.lsst.io/.

This paper is based on data collected at the Subaru Telescope and retrieved from the HSC data archive system, which is operated by the Subaru Telescope and Astronomy Data Centre (ADC) at NAOJ. Data analysis was in part carried out with the cooperation of Centre for Computational Astrophysics (CfCA), NAOJ. We are honoured and grateful for the opportunity of observing the Universe from Maunakea, which has the cultural, historical, and natural significance in Hawaii. 


\section*{Data Availability}
The data underlying this article will be shared on reasonable request to the corresponding author.



\bibliographystyle{mnras}
\bibliography{bibliography} 



\section*{Appendix A}
\label{sec:appendix}

\begin{table*}
\caption{ Sub-sample of galaxies with tidal structures. } 
\centering
\begin{tabular} {lllllll}  \hline\hline
        Name & R.A. & Dec. & PA & Structure & $q$ & $C_0$ \\ 
        & $\degree$ & $\degree$ & $\degree$ & & &  \\ \hline
        ES\_0.756\_-0.317 & 0.75613 & -0.31716 & 1.6 & 6 &  0.22 & -0.19 \\
        ES\_1.870\_0.902 & 1.87010 & 0.90169 & 24.0 & 6 &  0.35 & -0.65 \\
        ES\_2.265\_-0.583$^\star$ & 2.26488 & -0.58253 & -32.2 & 1 &  0.25 & 1.54 \\
        ES\_7.350\_0.316 & 7.35044 & 0.31575 & 22.6 & 5 &  0.21 & -0.52 \\
        ES\_13.728\_0.434 & 13.72821 & 0.43372 & 77.9 & 5 &  0.18 & 0.93 \\
        ES\_14.082\_-0.126 & 14.08171 & -0.12576 & -14.1 & 6 &  0.29 & 0.13 \\
        ES\_15.846\_-0.469 & 15.84552 & -0.46923 & -82.9 & 2 &  0.30 & 0.28 \\
        ES\_17.089\_0.044 & 17.08944 & 0.04368 & -14.8 & 1 &  0.28 & -0.15 \\
        ES\_18.239\_-0.345$^\star$ & 18.23926 & -0.34500 & -73.6 & 1, 2 &  0.31 & -0.10 \\
        ES\_18.624\_0.215 & 18.62440 & 0.21522 & 53.0 & 1 &  0.25 & 0.09 \\
        ES\_19.434\_0.254 & 19.43399 & 0.25434 & -64.7 & 2 &  0.43 & 0.05 \\
        ES\_19.677\_1.135 & 19.67695 & 1.13460 & -69.2 & 6 &  0.29 & 0.6 \\
        ES\_20.130\_0.786 & 20.13038 & 0.78572 & -28.7 & 5 &  0.20 & -0.72 \\
        ES\_20.271\_0.103 & 20.27080 & 0.10292 & 33.0 & 1 &  0.22 & 0.62 \\
        ES\_21.171\_0.081 & 21.17076 & 0.08109 & -50.6 & 3, 6 &  0.15 & 0.83 \\
        ES\_23.504\_-0.537 & 23.50383 & -0.53683 & -73.8 & 4 &  0.30 & -0.16 \\
        ES\_24.765\_0.437 & 24.76539 & 0.43667 & -20.7 & 5 &  0.21 & -0.15 \\
        ES\_28.265\_1.023 & 28.26460 & 1.02305 & 75.0 & 1 &  0.28 & 0.64 \\
        ES\_29.743\_-0.490 & 29.74327 & -0.48984 & -38.9 & 3 &  0.51 & -0.94 \\
        ES\_30.191\_-0.908 & 30.19102 & -0.90786 & 57.3 & 3 &  0.18 & 0.37 \\
        ES\_30.259\_0.031 & 30.25862 & 0.03129 & 50.4 & 3 &  0.16 & 0.46 \\
        ES\_31.960\_-0.384 & 31.96020 & -0.38421 & 2.8 & 1 &  0.28 & -0.51 \\
        ES\_31.972\_-0.028 & 31.97232 & -0.02835 & -34.6 & 1 &  0.31 & -0.58 \\
        ES\_32.794\_-0.517 & 32.79416 & -0.51687 & -52.2 & 1 &  0.22 & -0.11 \\
        ES\_37.936\_-0.945$^\star$ & 37.93556 & -0.94535 & -48.6 & 2 &  0.38 & -0.49 \\
        ES\_38.657\_-0.980 & 38.65716 & -0.97985 & -36.1 & 4 &  0.31 & -0.15 \\
        ES\_41.525\_0.564 & 41.52514 & 0.56355 & 46.7 & 6 &  0.28 & -0.25 \\
        ES\_43.963\_0.564 & 43.96324 & 0.56408 & 39.5 & 3 &  0.30 & -0.37 \\
        ES\_47.102\_-0.559 & 47.10241 & -0.55866 & 14.0 & 5 &  0.15 & -0.72 \\
        ES\_49.360\_-0.094$^\star$ & 49.36040 & -0.09421 & 117.0 & 5 &  0.28 & 0.39 \\
        ES\_52.605\_0.274 & 52.60547 & 0.27378 & -90.4 & 3 &  0.40 & -0.56 \\
        ES\_53.382\_0.699$^\star$ & 53.38157 & 0.69928 & 14.5 & 4 &  0.33 & 0.40 \\
        ES\_311.043\_-0.415 & 311.04300 & -0.41547 & 327.0 & 1 &  0.20 & -0.57 \\
        ES\_314.206\_-0.238$^\star$ & 314.20637 & -0.23774 & -5.2 & 1 &  0.29 & -0.15 \\
        ES\_315.137\_0.294 & 315.13693 & 0.29442 & 16.4 & 4 &  0.26 & -0.65 \\
        ES\_316.508\_-0.504$^\star$ & 316.50807 & -0.50361 & 36.9 & 4 &  0.32 & -0.62 \\
        ES\_319.092\_-0.730 & 319.09166 & -0.73014 & -35.1 & 1 &  0.17 & 0.10 \\
        ES\_323.211\_1.153 & 323.21103 & 1.15324 & 15.8 & 6 &  0.31 & -0.43 \\
        ES\_325.543\_-0.288 & 325.54260 & -0.28770 & 65.3 & 1,3 &  0.44 & 0.02 \\
        ES\_327.676\_0.912$^\star$ & 327.67642 & 0.91201 & -59.1 & 6 &  0.52 & 0.14 \\
        ES\_330.520\_-0.382 & 330.52009 & -0.38229 & -40.6 & 4 &  0.25 & -0.23 \\
        ES\_336.825\_-0.682 & 336.82471 & -0.68212 & 4.3 & 1 &  0.24 & 0.54 \\
        ES\_337.223\_0.771 & 337.22315 & 0.77054 & -27.3 & 2 &  0.35 & -0.72 \\
        ES\_339.889\_0.411 & 339.88871 & 0.41084 & -82.9 & 4 &  0.25 & -0.03 \\
        ES\_344.086\_0.487 & 344.08555 & 0.48698 & 82.8 & 4 &  0.25 & -0.24 \\
        ES\_344.174\_0.162 & 344.17380 & 0.16169 & 99.2 & 3 &  0.36 & -0.33 \\
        ES\_350.008\_-0.102 & 350.00776 & -0.10197 & 38.6 & 1 &  0.26 & -0.40 \\
        ES\_350.832\_-0.500 & 350.83207 & -0.49982 & -5.6 & 3 &  0.19 & -0.19 \\
        ES\_353.011\_1.223 & 353.01133 & 1.22329 & 13.5 & 5 &  0.24 & -0.17 \\
        ES\_354.826\_-0.643 & 354.82611 & -0.64319 & 80.5 & 2 &  0.29 & -0.44 \\ \hline\\[-0.5ex]
\end{tabular}
     \parbox[t]{170mm}{
    \textbf{Notes:}
Parameters of the galaxies in the table are Galaxy name, Right ascension (J2000), Declination (J2000), Position angle, the type of structures in accordance with the classification (1 -- tidal tails, 2 -- diffuse shells, 3 -- bridges, 4 -- arcs, 5  -- satellites or satellite remnants, 6 -- disk deformations), the apparent axis ratio of the modelling disk, the model disk boxyness/diskyness parameter. A star symbol $^\star$ marks a galaxy which is contained within or near the Milky Way's cirrus cloud or filaments.}\\

\label{table:table1}
 
\end{table*}

\begin{table*}
\caption{ Sub-sample of galaxies with structural features. } 
\centering
\begin{tabular} {lllllll}  \hline \hline
        Name & R.A. & Dec. & PA & Structure & $q$ & $C_0$ \\ 
        & $\degree$ & $\degree$ & $\degree$ & & &  \\ \hline
            ES82\_1.870\_0.902 & 1.8701 & 0.90169 & 24 & 1 & 0.35 & -0.66 \\ 
        ES82\_5.724\_-0.997 & 5.72384 & -0.99712 & -53 & 1, 2 & 0.49 & -0.65 \\ 
        ES82\_6.036\_0.379 & 6.03629 & 0.37913 & 61.3 & 1 & 0.25 & -0.97 \\ 
        ES82\_10.574\_0.371 & 10.57439 & 0.3707 & 63.16 & 2 & 0.28 & -0.39 \\ 
        ES82\_13.396\_0.596 & 13.3955 & 0.59626 & -43.9 & 2 & 0.38 & -0.55 \\ 
        ES82\_16.599\_-0.170 & 16.59915 & -0.16965 & 43.1 & 2 & 0.33 & -0.76 \\ 
        ES82\_17.693\_0.793 & 17.69276 & 0.79335 & 9.7 & 2 & 0.24 & -0.51 \\ 
        ES82\_17.898\_0.462 & 17.898 & 0.46228 & 88.29 & 2 & 0.19 & 0.07 \\ 
        ES82\_19.468\_0.781 & 19.46848 & 0.78079 & 3.7 & 2 & 0.28 & -0.4 \\ 
        ES82\_22.977\_0.492 & 22.97732 & 0.49202 & 15.5 & 2 & 0.17 & -0.34 \\ 
        ES82\_23.639\_-0.430 & 23.6386 & -0.4303 & 40.4 & 2 & 0.3 & -0.89 \\ 
        ES82\_23.752\_-0.451 & 23.75164 & -0.45117 & 14.7 & 2 & 0.33 & -0.56 \\ 
        ES82\_24.469\_1.043 & 24.46947 & 1.04303 & -57.4 & 2 & 0.32 & -0.59 \\ 
        ES82\_25.149\_0.259 & 25.14949 & 0.25854 & 45.96 & 1 & 0.16 & 0 \\ 
        ES82\_27.115\_0.394 & 27.11468 & 0.39393 & 2 & 2 & 0.41 & -0.79 \\ 
        ES82\_27.561\_0.062 & 27.56087 & 0.06211 & 21.6 & 2 & 0.24 & -0.46 \\ 
        ES82\_28.265\_1.023 & 28.2646 & 1.02305 & 74.99 & 1 & 0.28 & -0.04 \\ 
        ES82\_28.698\_-0.724 & 28.69831 & -0.7235 & -49.7 & 2 & 0.42 & -0.64 \\ 
        ES82\_29.237\_1.168 & 29.23671 & 1.16805 & 54.4 & 2 & 0.31 & -0.3 \\ 
        ES82\_29.743\_-0.490 & 29.74327 & -0.48984 & -38.9 & 3 & 0.51 & -0.94 \\ 
        ES82\_30.181\_1.019 & 30.18109 & 1.01942 & 15.7 & 2 & 0.25 & -1.13 \\ 
        ES82\_30.569\_-0.533 & 30.56906 & -0.53304 & 64.4 & 2 & 0.15 & -0.41 \\ 
        ES82\_31.245\_-0.264 & 31.24546 & -0.26359 & 89 & 2 & 0.17 & -0.78 \\ 
        ES82\_32.722\_-0.921 & 32.72181 & -0.92063 & 2.44 & 2 & 0.11 & 0.44 \\ 
        ES82\_32.777\_-0.655 & 32.77656 & -0.65528 & 17.38 & 1, 2 & 0.21 & 0.04 \\ 
        ES82\_33.789\_-0.704 & 33.78917 & -0.70357 & -26 & 2 & 0.15 & -0.69 \\ 
        ES82\_36.641\_-1.108 & 36.64127 & -1.10789 & 49.1 & 1 & 0.25 & 0.12 \\ 
        ES82\_37.911\_-1.083 & 37.9107 & -1.08262 & 84.17 & 2 & 0.33 & -0.42 \\ 
        ES82\_39.177\_-0.017 & 39.17708 & -0.01708 & 70.1 & 2 & 0.28 & -0.67 \\ 
        ES82\_40.079\_0.440 & 40.0789 & 0.43967 & 22 & 1 & 0.22 & -0.02 \\ 
        ES82\_42.143\_-0.993 & 42.14331 & -0.99324 & -1.5 & 2 & 0.3 & -0.93 \\ 
        ES82\_46.894\_-1.049 & 46.89375 & -1.04942 & -21.9 & 2 & 0.45 & -0.4 \\ 
        ES82\_49.642\_-0.362 & 49.64189 & -0.36224 & 25.6 & 2 & 0.38 & -0.85 \\ 
        ES82\_56.608\_-0.599 & 56.60797 & -0.59911 & -14.7 & 2 & 0.26 & -1.04 \\ 
        ES82\_313.353\_0.652 & 313.3526 & 0.65198 & 18.69 & 2 & 0.19 & -0.55 \\ 
        ES82\_314.393\_0.341 & 314.39313 & 0.34109 & -84.1 & 2 & 0.22 & -0.88 \\ 
        ES82\_319.276\_-0.627 & 319.276 & -0.62669 & 76 & 1 & 0.2 & -0.8 \\ 
        ES82\_320.078\_-1.093 & 320.07803 & -1.09339 & 81.5 & 1 & 0.4 & -0.48 \\ 
        ES82\_320.151\_1.212 & 320.15142 & 1.21234 & 32.52 & 2 & 0.3311 & -0.58 \\ 
        ES82\_326.305\_-0.186 & 326.30545 & -0.18557 & -84.68 & 2 & 0.29 & 0.17 \\ 
        ES82\_326.580\_-0.208 & 326.57991 & -0.20788 & 29.5 & 3 & 0.39 & -0.77 \\ 
        ES82\_331.378\_0.077 & 331.37818 & 0.07724 & 25.59 & 2 & 0.31 & -0.75 \\ 
        ES82\_333.748\_1.020 & 333.7481 & 1.01956 & -17.15 & 3 & 0.51 & -0.67 \\ 
        ES82\_336.167\_-0.897 & 336.16729 & -0.89715 & -23.1 & 2 & 0.42 & -0.59 \\ 
        ES82\_336.384\_0.039 & 336.38395 & 0.03903 & -5.7 & 2 & 0.35 & -0.52 \\ 
        ES82\_339.008\_-0.560 & 339.0076 & -0.56015 & 68.3 & 2 & 0.26 & -0.43 \\ 
        ES82\_340.030\_-0.818 & 340.03024 & -0.81837 & -33.8 & 2 & 0.29 & -0.01 \\ 
        ES82\_341.096\_-0.114 & 341.09551 & -0.11431 & -0.1 & 2 & 0.21 & -0.44 \\ 
        ES82\_346.594\_-0.822 & 346.59386 & -0.82193 & 16.26 & 2 & 0.12 & -0.09 \\ 
        ES82\_347.324\_0.064 & 347.32405 & 0.06416 & -55.3 & 2 & 0.23 & -0.86 \\ 
        ES82\_347.624\_0.273 & 347.62387 & 0.27306 & 84.5 & 2 & 0.27 & -0.38 \\ 
        ES82\_349.019\_-0.123 & 349.01868 & -0.12261 & 7.99 & 2 & 0.13 & 0.05 \\ 
        ES82\_351.140\_-0.731 & 351.13954 & -0.73058 & -30.2 & 2 & 0.36 & -0.42 \\ 
        ES82\_352.972\_-0.827 & 352.97156 & -0.8265 & 108 & 2 & 0.16 & -0.62 \\ 
        ES82\_353.011\_1.223 & 353.01133 & 1.22329 & 13.5 & 1 & 0.25 & -0.41 \\ 
        ES82\_355.229\_-1.074 & 355.22855 & -1.07404 & 33.7 & 2 & 0.3 & -0.56 \\ \hline \\
\end{tabular}
     \parbox[t]{170mm}{
    \textbf{Notes:}
Parameters of the galaxies in the table: Galaxy name, Right ascension (J2000), Declination (J2000), Position angle, The type of structures in accordance with the classification of structures given in the chapter of the same name (1 - lopsided discs , 2 -  disks with warps, 3 - polar rings), the apparent axis ratio of the modelling disk, the model disk boxyness/diskyness.}\\

\label{table:table2}
\end{table*}

\begin{table*}
\caption{ The structural parameters of edge-on galaxies from ES82. } 
\centering
    \begin{tabular}{lllllllllllll} \hline\hline
        Name & R.A. & Dec. & PA & $z$ & $M_r$ & Structure & Tidal & is\_edge\_on & isoA\_r & isoB\_r & $q$ & $C_0$ \\
        & $\degree$ & $\degree$ & $\degree$ & & mag & & & & arcsec & arcsec &  \\ \hline
        ES82\_0.048\_-0.644 & 0.04837 & -0.64402 & -62.7 & 0.14 & -22.094 & False & False & 2 & 10.29 & 3.48 & 0.25 & -0.1 \\
        ES82\_0.418\_1.092 & 0.41752 & 1.09196 & 18.6 & 0.061 & -21.451 & False & False & 4 & 18.54 & 9.19 & 0.49 & -0.35 \\ 
        ES82\_0.679\_-0.409 & 0.67949 & -0.40948 & -51.5 & 0.084 & -20.522 & False & False & 2 & 16.91 & 2.88 & 0.17 & -0.1 \\ 
        ES82\_0.722\_0.947 & 0.72167 & 0.94712 & -66.88 & 0.054 & -19.47 & False & False & 2 & 18.48 & 2.8 & 0.2 & 0.14 \\ 
        ES82\_0.756\_-0.317 & 0.75613 & -0.31716 & 1.6 & 0.084 & -21.493 & True & True & 1 & 16.91 & 4.43 & 0.25 & -0.19 \\ 
        ES82\_0.836\_-0.527 & 0.83576 & -0.52669 & -8 & 0.076 & -20.559 & False & False & 4 & 7.45 & 5.04 & 0.53 & -0.51 \\ 
        ... &  &  &  &  &  &  &  &  &  &  &  & \\
    \hline
    \end{tabular}
     \parbox[t]{170mm}{
    \textbf{Notes:}
Parameters of galaxies from the catalog ES 82: Galaxy id in the catalog, R.A. (J2000) in decimal degrees, Dec. (J2000) in decimal degrees, position angle, SDSS redshift, absolute magnitude in the $r$ band, boolean parameter characterizing the presence of any distinctive structures (see text) in the galaxy, boolean parameter characterizing the presence of tidal structures (see text) in the galaxy, galaxy group (1--4, see text), SDSS isophotal major axis in the $r$ band in arcseconds, SDSS isophotal minor axis in the $r$ band in arcseconds, disk apparent axis ratio, boxyness/diskyness parameter. 
Table is published in its entirety in the electronic edition.}\\ 
\label{table:table3}
\end{table*}

\FloatBarrier
\begin{figure*}
\centering
\includegraphics[width=18cm]{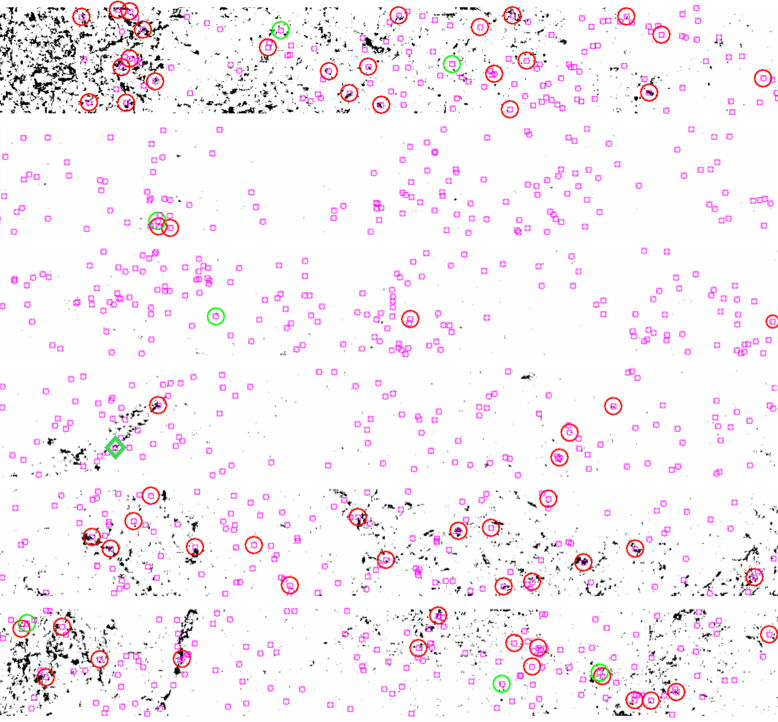}
\caption{A series of six stacked strips showing the Milky Way's cirrus clouds where the purple boxed circles show where each galaxy within our catalogue is located. Each strip, going from left to right, covers approximately 18 degrees of the sky. This covers a total of 110 degrees in R.A. The R.A. ranges from 60$\degree$ to 0$\degree$ which includes the first three strips and then 360$\degree$ to 310$\degree$ which includes the last three strips. The Dec. ranges from -1.25$\degree$ to 1.25$\degree$. These images are constructed from a single image (available on \url{https://physics.byu.edu/faculty/mosenkov/docs/cirrus_wcs.fits}) that spans the entire SDSS Stripe\,82 \citep{2023MNRAS.519.4735S}. 64 (red circles) out of 838 galaxies fall within or near a cirrus cloud. The seven green circles represent the galaxies that have tidal features. The one green diamond represents the galaxy that was removed as a tidal feature candidate.}
\label{fig:cirrus}
\end{figure*}
\FloatBarrier


\bsp	
\label{lastpage}
\end{document}